\documentclass[%
 reprint,
nofootinbib,
 amsmath,amssymb,
 aps,
]{revtex4-2}

\usepackage{graphicx}
\usepackage{dcolumn}
\usepackage{bm}

\usepackage{amsmath,amssymb,amsbsy,amstext,amsthm,simplewick,amsfonts}
\usepackage{mathrsfs}
\usepackage{graphicx}
\usepackage{wrapfig}
\usepackage{upgreek}
\usepackage{bm} 
\usepackage{framed}
\usepackage{bbm}
\usepackage{textcomp}
\usepackage{adjustbox}
\usepackage{makecell}
\usepackage{tcolorbox}
\usepackage{empheq}
\usepackage[normalem]{ulem}
\usepackage{enumitem}
\usepackage{braket} 
\usepackage{array}
\usepackage{dsfont}
\usepackage{physics}
\usepackage{ulem}
\usepackage{xcolor}
\usepackage{caption}
\usepackage{subcaption}
\usepackage{tocloft}
\usepackage{tikz}
\usepackage{xcolor}
\usetikzlibrary{patterns}

\usepackage{hyperref}
\hypersetup{colorlinks=true,
	breaklinks=true,
	pdfstartview=Fit,
	linkcolor=blue2,
	citecolor=red2,
	urlcolor=blue
}

\definecolor{red2}{RGB}{214, 39, 40}
\definecolor{green2}{RGB}{0,170,0}
\definecolor{blue2}{RGB}{0,100,200}
\definecolor{magenta2}{RGB}{191,64,191}
\definecolor{purple2}{RGB}{112,48,160}
\definecolor{orange2}{RGB}{255,192,0}

\newcommand{\LdS}{L_{\text{dS}}}
\newcommand{\LAdS}{L_{\text{AdS}}}
\newcommand{\mdreg}{$m$\&$d$ reg }

\graphicspath{{Fig/}}

\def\d{\mathrm{d}}

\def\Re{\mathrm{Re}\,}
\def\Im{\mathrm{Im}\,}

\newcommand{\vp}{\mathbf{p}}
\newcommand{\vq}{\mathbf{q}}
\newcommand{\vk}{\mathbf{k}}
\newcommand{\del}{\partial}

\usetikzlibrary{decorations.pathmorphing, decorations.markings,
                calc, positioning, arrows.meta, shapes.geometric}
\tikzset{
  prop/.style    = {thick},
  momentum/.style    = {thick, dotted},
  vertex/.style  = {circle, fill=black, inner sep=0pt, minimum size=2pt},
  loopvertex/.style  = {circle, draw=black, inner sep=0pt, minimum size=5pt},
}

\begin{document}

\preprint{APS/123-QED}


\title{All-Loop Renormalization and the Phase of the de Sitter Wavefunction }

\author{Alexander Farren}%
\affiliation{%
Department of Applied Mathematics and Theoretical Physics, University of Cambridge,\\Wilberforce Road, Cambridge, CB3 0WA, UK
}%

\author{Ciaran McCulloch}%
\affiliation{%
Department of Applied Mathematics and Theoretical Physics, University of Cambridge,\\Wilberforce Road, Cambridge, CB3 0WA, UK
}%

\author{Enrico Pajer}%
\affiliation{%
Department of Applied Mathematics and Theoretical Physics, University of Cambridge,\\Wilberforce Road, Cambridge, CB3 0WA, UK
}%

\author{Xi Tong}%
\affiliation{%
Department of Applied Mathematics and Theoretical Physics, University of Cambridge,\\Wilberforce Road, Cambridge, CB3 0WA, UK
}%


\begin{abstract}

    Cosmological observables of the primordial universe are encoded in the late-time field-theoretic wavefunction. For shift-symmetric scalars in de Sitter, a good approximation for many inflationary models, the wavefunction must be purely real at tree-level. This property is violated by a quantum anomaly in the process of renormalization. As a result, we show that the imaginary part of the wavefunction is fixed by its dependence on the renormalization scale to all loop orders in perturbation theory. This follows from unitarity, locality, dilation isometry and a Bunch-Davies state. The compact relation we uncover for the wavefunction implies an infinite set of relations among correlators of massless fields and their conjugate momenta, which we exemplify at one-loop order.
\end{abstract}

 \maketitle

\section{Introduction}

Ultraviolet (UV) contributions from loops in quantum field theory reflect our ignorance of the underlying high‑energy completion. The renormalization of these contributions generates a flow in theory space with a sliding scale. One may naively expect that in curved spacetimes these UV phenomena are a trivial re-writing of flat-space results because at very short distances spacetime curvature becomes negligible. However, this is not the case for at least two reasons. First, subleading UV divergences involve spacetime curvature terms and hence new operators beyond flat space. This is  made evident for example by the Weyl anomaly, when one recalls that the renormalization group (RG) can be thought of as a particular Weyl transformation (see e.g. \cite{localCS}). Second, the natural observables of flat and curved spacetime are crucially different. Hence, the renormalization scale becomes intertwined with kinematics in novel, interesting ways, as we discuss in the following. \\

A general feature of renormalization is its contribution to a quantum anomaly in scale transformations, often with dramatic consequences. For example, parity-odd correlators of massless scalars in de Sitter (dS) spacetime must vanish at tree-level by a powerful combination of unitarity, locality, dilation isometry and Bunch-Davies vacuum \cite{Liu:2019fag,Cabass:2022rhr,Goodhew:2024eup,Thavanesan:2025kyc}. However, at one loop, renormalization generates an imaginary part of the dS wavefunction, which in turn is directly observable in parity-odd correlators. This was shown in an explicit example in \cite{Lee:2023jby} and understood more generally as a consequence of Weyl anomaly \cite{Goodhew:2024eup} and renormalization \cite{Jain:2025maa}. A remarkable finding of \cite{Jain:2025maa} was that given any regulators compatible with unitarity, the imaginary part at one-loop is tied to the logarithmic renormalization scale dependence, and is therefore universal. This suggest deeper implications for the RG structure of cosmological wavefunctions, which we begin to uncover in this work.\\

Theoretical \cite{Tsamis:1996qq,Weinberg:2005vy,Marolf:2010zp,Krotov:2010ma,Bros:2006gs,Senatore:2009cf,Marolf:2010zp,Marolf:2010nz,Jatkar:2011ju,Pimentel:2012tw,Green:2020txs,Cohen:2020php,DiPietro:2023inn,Lee:2023jby,Heckelbacher:2022hbq,Chowdhury:2023khl,Green:2024fsz,Huenupi:2024ztu,Huenupi:2024ksc,Palma:2025oux,Benincasa:2024ptf,Braglia:2025cee,Nowinski:2025cvw,Bhowmick:2024kld,Bhowmick:2025mxh, Chowdhury:2023arc,Chowdhury:2024snc,Premkumar:2021mlz,Melville:2021lst,Cespedes:2023aal,Qin:2023bjk,Qin:2023nhv,Qin:2024gtr,Liu:2024xyi,Baumann:2024mvm,Jain:2025maa,Ansari:2025nng,Pimentel:2026kqc,Grafe:2026qsm,Chowdhury:2026upp} and phenomenological \cite{Chen:2016nrs,Wang:2021qez,Xianyu:2022jwk,Bodas:2025wuk,Kristiano:2023scm,Firouzjahi:2023ahg,Beneke:2023wmt,Fumagalli:2023zzl,Choudhury:2023rks,Firouzjahi:2023aum,Frolovsky:2025qre,Inomata:2025bqw,Fang:2025vhi,Inomata:2025pqa,Braglia:2025qrb,Ema:2025ftj,Wang:2025qfh,Maru:2021ezc,Niu:2022fki,Fujita:2023inz,Reinhard:2024evr,Garcia-Saenz:2025jis} explorations have sparked growing interest in loop effects in cosmology. Here we show that renormalization brings dramatic \textit{simplifications} in the structure of the wavefunction. Our result is summarised by a strikingly simple relation between the real and imaginary parts of the \textit{renormalized} dS wavefunction coefficients for $n$ massless fields,
\begin{align}
    \Im \widehat\psi_n =\tan\left(\frac{\pi}{2}\mu\partial_\mu\right) \Re \widehat\psi_n~,\label{tanEquation}
\end{align}
where $\mu$ denotes the renormalization scale of the theory. \eqref{tanEquation} is valid in momentum space to arbitrary loop orders for any theory of light fields with sufficiently soft interactions in the infrared (IR). It is a combined consequence of unitarity, locality, dilation isometry and the Bunch-Davies vacuum. This result generalizes the one-loop universality in \cite{Jain:2025maa} to all-loop orders and strengthens the connection to the RG flow. Since $\Im \widehat \psi_n$ appears in boundary correlators of fields and their conjugate momenta, \eqref{tanEquation} also implies an infinite set of relations among these observables. To derive \eqref{tanEquation} in perturbation theory, we show that to any finite $L$-loop order, the renormalized wavefunction coefficients in IR-finite theories take the universal form:
\begin{align}\label{psinform}
    \widehat \psi_n^{(L)}=\sum_{\ell=0}^L a_\ell \, \left(\log\frac{\mu}{H}+\frac{i\pi}{2}\right)^\ell~,
\end{align}
where $a_\ell=a_\ell(\{\mathbf{k}\};\{\lambda\},H,\cdots)$ are \textit{real} functions of kinematics that \textit{rationally}
depend on the couplings $\{\lambda\}$ and the (explicit) Hubble parameter $H$. 

The insightful reader will notice that \eqref{tanEquation}, while providing a powerful all-loop theoretical constraint, does not involve kinematics and so does not have obvious implications for observations. However, as in Minkowski, we expect the running with $\mu$ to be accompanied by running in kinematics $\{\mathbf{k}\}$, presumably partial energies appearing in dilation-invariant ratios. We hope to elucidate this in a future publication. Additionally, \eqref{tanEquation} echos the flat-space result for form factors \cite{Caron_Huot_2016} while \eqref{psinform} closely resembles the asymptotic expansion of on-shell amplitudes, where unitarity alone could shed light on the RG flow \cite{Chavda:2025aqm,Chavda:2025awr}.
Our results suggest an analogous architecture may exist for dS wavefunctions.

\section{Wavefunction coefficients to all loops}

For simplicity and concreteness, we first study an EFT of a derivatively coupled massless scalar in dS,
\begin{align}
    S=\int\d^4 x\sqrt{-g}\left[-\frac{1}{2}\left(\partial_\mu\phi\right)^2+\mathcal{L}_{\rm int}(\partial\phi,\partial^2\phi,\cdots)\right]~.\label{generalIRsafeScalarEFTLagrangian}
\end{align}
The interaction Lagrangian consists of an infinite tower of higher dimensional operators parametrised by
\begin{align}
    \mathcal{L}_{\rm int}=\sum_{p,q}\frac{\lambda_{p,q}}{\Lambda^{p-4}} \left[D_q\left(\frac{-H\eta\partial_\eta}{\Lambda},\frac{-H\eta\partial_i}{\Lambda}\right)\phi^p\right]~,\label{interactionLagrangianIn3d}
\end{align}
where $\Lambda$ is the cutoff and $D_q(\cdot,\cdot)$ represents a degree-$q$ real polynomial of derivatives acting on the fields. All spatial indices are contracted using $\delta_{ij}$ and $\epsilon_{ijk}$ to ensure rotational invariance. Dilation invariance dictates that the dimensionless couplings $\{\lambda\}$ must be independent of time while unitarity enforces their reality. Crucially, we require the number of derivatives $q$ in \eqref{interactionLagrangianIn3d} to be sufficiently large that there is no late-time/soft-momentum divergence in the theory. This therefore defines a class of IR-safe EFTs in dS.\footnote{We make the reasonable assumption that IR-safe interactions do not induce either late-time or soft-momentum divergences.}

The state of quantum fields at the conformal boundary $\eta_0\to 0$ is characterised by the field-theoretic wavefunctional,
\begin{align}
    \Psi[\varphi]=\int_{\phi(-\infty)=0}^{\phi(\eta_0)=\varphi}[\mathcal{D}\phi]\,e^{i S[\phi]}~,
\end{align}
which is computed as a path integral that starts with the Bunch-Davies vacuum at $\eta\to-\infty$ and ends with the prescribed boundary value $\varphi$ at $\eta_0\to 0$. The wavefunctional in momentum space is conventionally parametrised by 
\begin{align}
    \label{WFUDefinSect2}
    \nonumber\log \Psi&=+\frac{1}{2!}\int_{\mathbf{k}}\widehat \psi_2(k)\varphi_\mathbf{k} \varphi_{-\mathbf{k}}\\
    \nonumber&\quad +\frac{1}{3!}\int_{\mathbf{k}_1+\mathbf{k}_2+\mathbf{k}_3=0}\widehat \psi_3 (\mathbf{k}_1,\mathbf{k}_2,\mathbf{k}_3) \varphi_{\mathbf{k}_1}\varphi_{\mathbf{k}_2}\varphi_{\mathbf{k}_3}\\
    &\quad +\cdots~,
\end{align}
where the wavefunction coefficients $\widehat\psi_n(\{\mathbf{k}\})$ can be computed order-by-order in perturbation theory using diagrammatics. We denote the loop expansion by
\begin{align}
\widehat\psi_n=\psi_n^{(0)}+\widehat{\psi}_n^{(1)}+\widehat{\psi}_n^{(2)}+\cdots~
\end{align}
and use a hat to denote renormalized quantities. As in Minkowski QFTs, loop diagrams are generally UV-divergent and must be regularised and renormalized by counterterms. For instance, in dimensional regularization, one continues the spatial dimensions to $d=3-\epsilon$ and rewrites the interactions \eqref{interactionLagrangianIn3d} as
\begin{align}
     S_{\rm int}&=\int\d\eta\d^d x\sqrt{-g_{d+1}} \label{interactionLagrangianIndd}\\
    \nonumber&\quad\sum_{p,q}\frac{\lambda_{p,q}}{\Lambda^{p-4}}\left[D_q\left(\frac{-H\eta\partial_\eta}{\Lambda},\frac{-H\eta\partial_i}{\Lambda}\right)\phi^p\right]\mu^{(\frac{p}{2}-1)\epsilon}~,
\end{align}
where we have inserted appropriate powers of a renormalization scale $\mu$ to keep $\{\lambda\}$ dimensionless. The counterterms $S_{\rm int}^{\rm ct}$ induced in the EFT take the same form as \eqref{interactionLagrangianIndd} but with the couplings replaced by $\lambda_{m,p}\mapsto \delta_{m,p}$, where $\{\delta\}$ are real coefficients calibrated to cancel UV divergences in the loops. At $L$-loop order, the renormalized wavefunction requires the inclusion of all lower-loop diagrams involving counterterm-couplings,
\begin{align}
    \widehat{\psi}_n^{(L)}=\psi_n^{(L)}+\psi_n^{{\rm ct},(L-1)}+\cdots+\psi_n^{{\rm ct},(0)}~.\label{fromPsiToPsiHat}
\end{align}
The removal of divergences for all $\widehat{\psi}_n$ determines $\{\delta\}$ order by order in loops. We now prove \eqref{psinform} using simple power-counting arguments in two versions of dimensional regularization.

\vskip 7pt
{\bf \noindent Usual dimensional regularization.}\quad In this scheme, one continues the dimensionality alone holding the scalar mass fixed at $m^2_\phi=0$. For an $L$-loop diagram with $I$ internal lines and $V$ vertices, the wavefunction coefficient is schematically computed by
\begin{align}
    \psi_n^{(L)}=\int_{\mathbf{q}_1\cdots\mathbf{q}_L}\int_{-\infty}^0\left[\,\prod_{v=1}^V   \d\eta_v V_v\right]\left[\,\prod_{e=1}^n K_e\right] \left[\,\prod_{{\rm i}=1}^I G_{\rm i}\right]~,\label{LloopPsiGeneral}
\end{align}
where $K_e,G_{\rm i}$ are bulk-boundary and bulk-bulk propagators and
\begin{align}
    V_v\equiv \frac{\mu^{\left(\frac{p_v}{2}-1\right)\epsilon}}{(-H\eta_v)^{4-\epsilon}} \frac{i\lambda_v}{\Lambda^{p_v-4}}\, D_v\left(\frac{-H\eta_v\partial_{\eta_v}}{\Lambda},\frac{-i H\eta_v\mathbf{k}_i}{\Lambda}\right)
\end{align}
denote $p_v$-point vertex operators. The propagators are constructed from the free theory in $d$ dimensions and involve products of Bessel functions. These diagrammatic building blocks satisfy simple properties under a Wick rotation\footnote{The Wick rotation is justified by the Bunch-Davies $i\varepsilon$-prescription, which exponentially damps out the contribution from the arc at $|\eta|\to\infty$, $\Im \eta>0$.} $\eta=iz$, $z>0$, \cite{Jain:2025maa}
\begin{subequations}\label{dimregKGWickRules}
    \begin{align}
        K(i z,k)&= \text{real}~,\\
        G(i z_1,i z_2;k)&=i^{\epsilon} H^{-\epsilon}\times \text{real}~,
    \end{align}
\end{subequations}
where we have also extracted the non-integer-power dependence in the Hubble. Vertices transform as
\begin{align}
    \d\eta_v V_v=i^{-\epsilon}\mu^{\left(\frac{p_v}{2}-1\right)\epsilon} H^\epsilon \times \text{real}~.\label{dimregVWickRules}
\end{align}
Collecting \eqref{dimregKGWickRules}-\eqref{dimregVWickRules} and using the topological identities $L=I-V+1$ and $\sum_v p_v=2I+n$, we can strip all $\epsilon$-dependent powers of $i$, $\mu$ and $H$ as
\begin{align}
    \psi_n^{(L)}\equiv (i\mu/H)^{(L-1)\epsilon}\mu^{n\epsilon/2} \times Q_n^{(L)}~,\label{dimregpsinLloopPowerExtracted}
\end{align}
with $Q_n^{(L)}$ being a loop integral that is real, $\mu$-independent, and rational in $H$. Note that the $i$-factors reproduce the phase formula of \cite{Goodhew:2024eup}. We can then formally write this $L$-loop integral as an $\epsilon$-expansion,
\begin{align}
    Q_n^{(L)}=\frac{f_L^{(L)}}{\epsilon^L}+\cdots+\frac{f_1^{(L)}}{\epsilon}+f_0^{(L)}+\mathcal{O}(\epsilon)~,\label{dimregLloopPsiRealIntegralExpansion}
\end{align}
The residues $f_j^{(L)}$ naturally inherit the reality, $\mu$-independence, and $H$-rationality of $Q_n^{(L)}$. Note that the bare wavefunction coefficient is \textit{not} a uniform function of $\log(i\mu/H)$ after the $\epsilon$-expansion. 

Structure emerges when we look at \textit{renormalized} quantities, using \eqref{fromPsiToPsiHat}. As noted above, the mathematical form of $\ell$-loop counterterm diagrams $\psi_n^{{\rm ct},(\ell)}$ is identical to that of the $\ell$-loop bare diagram, albeit with $\{\lambda\}$ partially replaced by $\{\delta\}$. We can thus construct a quantity
\begin{align}
     \nonumber\widehat{Q}_n(\epsilon)&\equiv (i\mu/H)^{-(L-1)\epsilon}\mu^{-n\epsilon/2}\times\widehat\psi_n^{(L)}(\epsilon)\\
    &=\sum_{j=0}^L \frac{f^{(L)}_j}{\epsilon^j}+\sum_{\ell=0}^{L-1}\sum_{j=0}^L \left(\frac{i\mu}{H}\right)^{(\ell-L)\epsilon}\frac{f^{{\rm ct},(\ell)}_j}{\epsilon^j}~,\label{QfunctionExpandedForm}
\end{align}
which \textit{is} a uniform function of $\log(i\mu/H)$ after expansion. Crucially, the cancellation of all $\epsilon$-poles in $\widehat\psi_n^{(L)}(\epsilon)$ implies regularity of $\widehat{Q}_n(\epsilon)$ at $\epsilon=0$, leaving the $\epsilon \to 0$ limit as
\begin{align}
    \widehat\psi_n^{(L)}(0)&=1\times \widehat{Q}_n(0)=\sum_{\ell=0}^L a_\ell \left(\log\frac{\mu}{H}+\frac{i\pi}{2}\right)^\ell~,\label{desiredForm}
\end{align}
with $a_\ell$ a real and linear combinations of $f^L_j$ and $f_j^{{\rm ct},(\ell)}$ (see Appendix \ref{appendixTwoloopExample}). This concludes the proof of \eqref{psinform} in dim reg.

\vskip 7pt
{\bf \noindent Mass \& dimensional regularization.} \quad The \mdreg is devised to reduce the computational difficulty in the usual dim reg, which necessarily involves Bessel functions in the propagators. In \mdreg, one simultaneously continues the mass of the field and dimensionality such that the combination $\nu\equiv \sqrt{d^2/4-m_\phi^2(d)/H^2}=3/2$ stays fixed. The propagators thus simplify to simple exponentials with the following Wick-rotation behaviour:
\begin{subequations}\label{mdregKGWickRules}
    \begin{align}
        K(i z,k)&= i^{\epsilon/2}(-\eta_0)^{\epsilon/2}\times \text{real}~,\\
        G(i z_1,i z_2;k)&=i^{\epsilon} H^{-\epsilon}\times \text{real}~,
    \end{align}
\end{subequations}
where we have also extracted the irrational dependence on the boundary time $\eta_0$. Eq. \eqref{dimregpsinLloopPowerExtracted} now becomes
\begin{align}
    \psi_n^{(L)}= (i\mu/H)^{(L-1+n/2)\epsilon}(-H\eta_0)^{n\epsilon/2}\times Q_n^{(L)}~,\label{mdregpsinLloopPowerExtracted}
\end{align}
with $Q_n^{(L)}$ being real, $\mu$-independent and $H$-rational. Thus the bare loop diagram is again not a uniform function of $\log(i\mu/H)$. Worse still, it contains spurious secular terms $\log(-\eta_0)$ at late times which should be physically absent in IR-safe theories. These problems are once again solved by renormalization, through the same argument as in \eqref{QfunctionExpandedForm}. The subtraction of UV divergences automatically brings the form of $\widehat\psi_n^{(L)}$ to \eqref{psinform} while removing spurious secular terms.

\vskip 7pt
{\bf \noindent Euclidean AdS.}\quad Calculations in dS can be related to Euclidean anti-de Sitter (EAdS), a.k.a. hyperbolic space, by a double Wick rotation: $\eta \to \pm i z$ and $ \LdS\to\pm i \LAdS$, where $z$ is the radial coordinate in the Poincar\'e patch of EAdS, and $L_\text{dS}\equiv 1/H$ and $\LAdS$ are the radii of dS and EAdS respectively \cite{Maldacena:2002vr}. Naively the metric only involves $\LdS^2$ or $\LAdS^2$ and the choice of sign seems irrelevant. However the action, and hence all interaction vertices, contain $\sqrt{-g}$ and so the sign distinction and the corresponding path in the complex plane matters. In most of the literature, the (EA)dS radius rotation has been chosen to be $\LdS\to + i \LAdS$ \cite{Maldacena:2002vr,Anninos:2014lwa,Sleight:2020obc}, with the exception of \cite{Harlow:2011ke}. The authors of \cite{Bzowski:2023nef} commented that this choice is inconsistent in even space dimension but fine in odd. This choice is inconsequential when only $L_\text{(EA)dS}^2$ appears (in fact \cite{McFadden:2009fg} didn't commit to a sign), but has dramatic consequences when UV-divergences are renormalized. Using this prescription on \eqref{psinform} we would find $\log(-\mu \LAdS)$ and hence a complex EAdS wavefunction, which violates unitarity. Here we show that the opposite choice, namely $\LdS=-i \LAdS$ allows for a path in complexified metrics that respects unitarity. Indeed consider the metric
\begin{align}
    ds^2=\frac{L(\theta)^2}{\xi(\theta)^2} \left[ e^{i(\pi-\epsilon)}d\xi(\theta)^2+d\mathbf{x}^2 \right]\,,
\end{align}
where $\theta\in \mathbb{R}$ is a real parameter and $0<\epsilon \ll 1$ is the Lorentzian $i\epsilon$ prescription. The choice $\LdS\to + i \LAdS$ is obtained from setting $L(\theta)=|L|e^{i\theta}$, $\xi(\theta)=|\xi|e^{-i\theta}$ and the path $\theta \in [0,\pi/2]$. In this way $\theta=0$ corresponds to dS and $\theta=\pi/2$ to EAdS. One finds that $|\arg(g_{00})|=\pi-\epsilon +2\theta$ and $|\arg(g_{ii})|=3\times 4\theta$ hence violating the Kontsevich-Segal-Witten criterion $\sum |g_{\mu\mu}|<\pi$ \cite{Kontsevich:2021dmb,Witten:2021nzp}. Conversely, the choice $\LdS\to - i \LAdS$ is obtained from setting $L(\theta)=|L|e^{-i\theta}$, $\xi(\theta)=|\xi|e^{-i\theta}$ and the path $\theta \in [0,\pi/2]$. While the end points are the same, now one finds $|\arg(g_{00})|=\pi-\epsilon -2\theta$ and $|\arg(g_{ii})|=0$, satisfying this criterion. Appling this rotation to our dS result we find as expected,
\begin{align}
    \text{EAdS: }\widehat\psi_n^{(L)}=\sum_{\ell=0}^L a_\ell \log^\ell(\mu \LAdS)\,,
\end{align}
which is real provided that $a_\ell$ only depends on even powers of $\LAdS$, in agreement with the phase formula of \cite{Goodhew:2024eup}.

\section{From the real to the imaginary}

To better illustrate the implication of \eqref{psinform}, we note that $\widehat\psi_n^{{(L)}}$ is a complex RG running away from being real,
\begin{align}
    \label{eq:mu_shift}
    e^{-\frac{i\pi}{2} \mu\partial_\mu}\widehat\psi_n^{(L)}=\sum_{\ell=0}^L a_\ell \log^\ell\frac{\mu}{H}=e^{+\frac{i\pi}{2} \mu\partial_\mu}\left[\widehat\psi_n^{(L)}\right]^*~.
\end{align}
Simple algebra then leads to \eqref{tanEquation}:
\begin{align}\nonumber
    \Im \widehat\psi_n =\frac{1}{2i}\left(1-e^{-i\pi\mu\partial_\mu}\right)\widehat\psi_n=\tan\left(\frac{\pi}{2}\mu\partial_\mu\right) \Re \widehat\psi_n~,
\end{align}
where we have dropped the loop label since it is valid up to any finite $L$ loops. As a consistency check, truncating to $L=1$ and using $\tan\left(\frac{\pi}{2}\mu\partial_\mu\right)\approx \frac{\pi}{2}\mu\partial_\mu$ reproduces the one-loop universality equation in \cite{Jain:2025maa}. Note how our assumptions contribute to the above result. Unitarity enforces the reality of the couplings/counterterm coefficients, while the dilation isometry constrains the form of the interactions. Locality and the Bunch-Davies vacuum allow the Wick rotation in the time integrals. IR finiteness guarantees a smooth late-time limit and the absence of soft/collinear singularities in the momentum integrals. Finally, assuming parity, one can use \eqref{WFUDefinSect2} to recast the above equation \eqref{tanEquation} as
\begin{equation}
   \Im\log\Psi= \tan(\frac{\pi}{2}\mu\partial_\mu)\Re\log\Psi~,
\end{equation}
which is a constraint on the phase of the wavefunction.

\vskip 7pt
{\bf \noindent Generalization.}\quad Our result generalizes to other theories, including: (i) any bulk fields whose masses lie in the complementary series, as their bulk-bulk propagators also obey \eqref{dimregKGWickRules}. (ii) any bosonic spinning fields with scale-invariant and IR-safe interactions, as then the $\eta$-power counting remains the same. (iii) boost-breaking effects in the free theory such as non-unit sound speeds and chemical potentials, as the Wick rotation is insensitive to these effects. In all such cases, the key ingredients of the proof remain unchanged and we once again arrive at \eqref{tanEquation}. One notable distinction, however, is that with massive bulk fields, the $H$-dependence in the coefficients of \eqref{psinform} is no longer rational, i.e. $a_\ell=a_\ell(\{\mathbf{k}\};\{\lambda\},\Lambda,H,\{\nu\})$ with $\nu=\sqrt{d^2/4-m^2/H^2}$ being the dimensionless mass of a complementary bulk field. This does not affect its power counting because the $\epsilon\to 0$ limit of $\nu$ is smooth without extra $\log H$ terms. It is also compatible with the EAdS perspective as $\nu$ always Wick-rotates to a positive real number.

\vskip 7pt
{\bf \noindent Comparison to amplitudes.}\quad For Minkowski amplitudes, the imaginary part is caused by the on-shell production of particles in the loop and is \textit{non-linearly} constrained by the optical theorem \cite{Chavda:2025aqm},
\begin{align}
    \Im \mathcal{A}_{n+m}= \frac{1}{2}\sum_{k\text{-cuts}}\int \d\Pi_k  \mathcal{A}_{n+k} \mathcal{A}^*_{m+k}~.
\end{align}
In comparison, \eqref{tanEquation} is \textit{linear}. The difference is due to the absence of kinematic thresholds in the physical domain (i.e. the absence of folded singularities), so that the only imaginary contribution originates from the regulated volume measure. In other words, one cannot adjust external momenta to produce on-shell particles by crossing the branch cuts. However, our result does bear a remarkable resemblance to Minkowski \textit{form factors}. When written as \eqref{eq:mu_shift}, the phase relation for renormalised wavefunctions echoes the finding of \cite{Caron_Huot_2016}, which equates the phase of the massless S-matrix to dilatations via form factors. Both results restrict the running of the phase of the wavefunction using unitarity.

\def\mathLarge#1{\mbox{\Large $#1$}}
\DeclareRobustCommand{\PPi}{\displaystyle \mathLarge{\mathLarge{\pi}}}


\section{Correlator relations}

Having established \eqref{tanEquation}, we now show that it translates to an infinite set of relations for correlators of the boundary field $\varphi$ and its conjugate momentum $\PPi$,
\begin{equation}
    \langle \mathcal{O}(\varphi,\PPi)\rangle =\frac{\int \mathcal{D}\varphi \, \Psi^*[\varphi] \,\mathcal{O}(\varphi,-i\frac{\delta}{\delta \varphi})\, \Psi[\varphi]}{\int \mathcal{D}\varphi\, \Psi^*[\varphi]\, \Psi[\varphi]}~.\label{BornRule}
\end{equation}
For simplicity, we assume parity invariance in this section. Denoting $\widehat\psi_n\equiv\widehat \rho_n/2+i \widehat\gamma_n$, \eqref{tanEquation} becomes
\begin{align}
    \tan(\frac{\pi}{2}\mu\partial_\mu)\widehat\rho_n=2\widehat\gamma_n~.\label{tanEquation3}
\end{align}
Let us focus on connected correlators. One can compute these perturbatively around the Gaussian wavefunction in terms of $\{\widehat\rho_n,\widehat\gamma_n\}$.

For example, consider the connected, renormalized correlators
\begin{subequations}
    \label{eq:defCorrelators}
    \begin{align}
    &\widehat B_2(k)\equiv \langle \varphi_{\mathbf{k}}\varphi_{-\mathbf{k}}\rangle'~,\qquad \widehat C_2(k)\equiv \langle \{\varphi_{\mathbf{k}},\PPi_{-\mathbf{k}}\}\rangle'~,\\
    &\widehat B_3(k_1,k_2,k_3)\equiv \langle \varphi_{\mathbf{k}_1}\varphi_{\mathbf{k}_2}\varphi_{\mathbf{k}_3}\rangle'~,\\
    &\widehat C_3(k_1,k_2;k_3)\equiv \langle \varphi_{\mathbf{k}_1}\varphi_{\mathbf{k}_2}\PPi_{\mathbf{k}_3}\rangle'~.
    \end{align}
\end{subequations}
where the prime means we stripped off the overall momentum-conserving delta function. Let us focus consider a cubic theory with coupling $\lambda$. Since $n$-point correlators scale as $\lambda^{n+2(L-1)}$, `zero-loop' refers to terms scaling as $\lambda^{n-2}$ while `one-loop' refers to $\lambda^n$. At zero-loop,
\begin{equation}
\widehat B_2^{(0)} =-\frac{1}{\widehat\rho_2^{(0)}},\qquad \widehat C_2^{(0)}=-\frac{\widehat\gamma_2^{(0)}}{\widehat\rho_2^{(0)}}~,
\end{equation}
leading to the familiar power spectrum. The late-time divergence \smash{$\widehat\gamma_2^{(0)}\sim i\eta_0^{-1}$} can be removed by adding a local counterterm on the boundary \cite{Maldacena:2002vr}, thereby setting $\widehat C_2^{(0)}\sim \widehat\gamma_2^{(0)}=0$. Truncating \eqref{tanEquation3} to zero loops further yields $\widehat C_n^{(0)}\sim \widehat\gamma_n^{(0)}=0$ for $n\geq 3$. With this prescription, we have
\begin{align}
    \widehat B_2^{(1)}=\frac{\widehat\rho_2^{(1)}}{\left(\widehat\rho_2^{(0)}\right)^2} +\mathrm{classical\ loops}~, 
    \quad \widehat C_2^{(1)}=-\frac{\widehat\gamma_2^{(1)}}{\widehat \rho_2^{(0)}}~.
\end{align}
Solving backwards yields the reconstruction
\begin{align}
    \widehat \rho_2^{(1)}=\frac{\widehat B_2^{(1)}}{\left(\widehat B_2^{(0)}\right)^2} +\mathrm{classical\ loops}~, 
    \quad \widehat \gamma_2^{(1)}=\frac{\widehat C_2^{(1)}}{\widehat B_2^{(0)}}~.
\end{align}
Insertion into the one-loop truncation of \eqref{tanEquation3} gives\footnote{We assume the UV convergence of classical loops thanks to IR safety, hence they are annihilated by $\mu\partial_\mu$.}
\begin{subequations}
    \begin{align}
    \mu\partial_\mu \widehat B_2^{(1)}(k)&=\frac{4}{\pi}\widehat C_2^{(1)}(k)\widehat B_2^{(0)}(k)~,\label{2ptConsistencyRelation}\\
    \mu\partial_\mu \widehat C_2^{(1)}(k)&=0~.
    \end{align}
\end{subequations}
Similarly, at three points we find (see Appendix \ref{appendix3ptConsistencyRelation}),
\begin{subequations}
    \begin{align}
    \mu\partial_\mu \widehat B_3^{(1)}(k_1,k_2,k_3)&=\frac{4}{\pi}\Bigg[\widehat C_3^{(1)}(k_1,k_2;k_3)\widehat B_2^{(0)}(k_3)\label{3ptConsistencyRelation}\\
    \nonumber+\widehat B_3^{(0)}(k_1,&k_2,k_3)\left(\widehat C_2^{(1)}(k_1)+ \widehat C_2^{(1)}(k_2)\right)\Bigg]~,\\
    \mu\partial_\mu \widehat C_3^{(1)}(k_1,k_2;k_3)&=0~.
    \end{align}
\end{subequations}

To summarise, \eqref{BornRule} maps out a dictionary that translates wavefunctions to correlators. Conversely, one can reconstruct the wavefunctions from correlators using the inverted dictionary \cite{Stefanyszyn:2024msm}. It is therefore clear that for each pair $\{\widehat\rho_n,\widehat\gamma_n\}$, one can obtain a relation for correlators $\{B,C\}$ using their reconstructions.

\section{Summary}
We derived a new constraint on the structure of the dS wavefunction, \eqref{tanEquation}, which holds at all loop orders in perturbation theory on a rigid dS background in the Bunch-Davies vacuum. The constraint requires local and unitary interactions, which are finite in the IR and late-time limits, involving massless external fields exchanging light fields in the complementary series of dS. The constraint arises because every logarithm of the renormalization scale $\mu$ is of the form $\log\left( i\mu/H \right)$, and implies a relation between boundary correlators of massless fields and their canonical momenta. In passing, we clarified the correct analytic continuation from dS to EAdS, addressing a discrepancy in the literature.

Since the renormalization scale $\mu$ is unobservable, the reader may question the utility of our results. One most immediate application is as a consistency check of loop calculations in dS. More significantly, the $\left( -H\eta \right)^{\epsilon}$ factors that led to the correspondence between terms in $\log i$ and $\log \mu$ also lead to logarithmic branch points in the total or partial energies of the external kinematics \cite{Salcedo:2022aal,Lee:2023jby,Agui-Salcedo:2023wlq,Bhowmick:2025mxh}. Consequently, we anticipate that the $\mu$-dependence of the wavefunction or correlators is also linked to a kinematic dependence. 

There may be a connection between our results and those in \cite{Chavda:2025aqm,Chavda:2025awr, Caron_Huot_2016}, where it is claimed that in some theories the Callan-Symanzik equation may be a direct consequence of unitarity of the $S$-matrix, independent of any expansion in Feynman diagrams. Our results also use unitarity to recursively constrain the renormalization scale-dependence of the wavefunction, and this can be made explicit using the cosmological optical theorem \cite{Goodhew:2020hob,Melville:2021lst,Goodhew:2021oqg}.

Finally, we believe our result can be understood as a consequence of a quantum anomaly of Weyl symmetry with factor $\Omega=e^{- i \pi}$. This was understood as a result of covariant \cite{Goodhew:2024eup} and unitary \cite{Jain:2025maa} renormalisation, but the consequence for the renormalised  $L$-loop wavefunctions was not fully clarified until our work. We hope to describe the role of the Weyl anomaly in cosmological observables in the future, perhaps employing a Callan-Symanzik-like equation. 

\vskip 10pt

\paragraph*{Acknowledgements} We thank Diksha Jain, Sebastian Mizera, Shota Komatsu, Aron Wall for helpful discussions and comments on the draft. X.T. thanks the Yukawa Institute for Theoretical Physics at Kyoto University for helpful discussions during ``Progress of Theoretical Bootstrap''. X.T. is supported by STFC consolidated grants ST/T000694/1 and ST/X000664/1. C.M.~is supported by STFC training grant ST/W507350/1. A.F. is supported by the Robert Gardiner and Helen Stone awards of the Cambridge Trust. For the purpose of open access, the authors have applied a CC BY public copyright licence to any Author Accepted Manuscript version arising.  



\newpage

\appendix

\onecolumngrid

\section{Renormalization and power counting: a two-loop example}\label{appendixTwoloopExample}

In this appendix, we compute $\widehat{\psi}_n^{(2)}$ in dim reg and explicitly show how the logarithms combine into the desired form \eqref{desiredForm} upon requiring the cancellation of divergences. Expanding at $\epsilon\to 0$ up to $L=2$, the bare loop and the counterterms read
\begin{subequations}
\begin{align}
    \nonumber \psi_n^{(2)}&=\frac{f_2^{\text{(2)}}}{\epsilon ^2}+\frac{\left(f_1^{\text{(2)}}+\left(\frac{i \pi}{2}-\log H \right)
    f_2^{\text{(2)}}\right)+\frac{n+2}{2} f_2^{\text{(2)}} \log \mu
   }{\epsilon
   }\\
   \nonumber& \qquad\quad~ +f_0^{\text{(2)}}+\frac{\pi +2 i \log H}{8} \left(4 i
   f_1^{\text{(2)}}-f_2^{\text{(2)}} (\pi +2 i \log H)\right)\\
   & \qquad\quad~ +\frac{n+2}{2} \left(f_1^{\text{(2)}}+\left(\frac{i \pi}{2}-\log H \right)
   f_2^{\text{(2)}}\right) \log \mu +\frac{(n+2)^2}{8}  f_2^{\text{(2)}} \log
   ^2\mu~,\\
   \nonumber \psi_n^{{\rm ct},(1)}&=\frac{f_2^{{\rm ct},\text{(1)}}}{\epsilon ^2}+\frac{f_1^{{\rm ct},\text{(1)}}+\frac{n}{2}
   f_2^{{\rm ct},\text{(1)}} \log \mu }{\epsilon
   }\\
   & \qquad\qquad\, +f_0^{{\rm ct},\text{(1)}}+\frac{n}{2} f_1^{{\rm ct},\text{(1)}} \log \mu +\frac{n^2}{8}
   f_2^{{\rm ct},\text{(1)}} \log ^2\mu ~,\\
   \nonumber \psi_n^{{\rm ct},(0)}&=\frac{f_2^{{\rm ct},\text{(0)}}}{\epsilon ^2}+\frac{\left(f_1^{{\rm ct},\text{(0)}}+\left(\log H-\frac{i \pi}{2} \right)
    f_2^{{\rm ct},\text{(0)}}\right)+\frac{n-2}{2} f_2^{{\rm ct},\text{(0)}} \log \mu
   }{\epsilon
   }\\
   \nonumber&\qquad\qquad\, + f_0^{{\rm ct},\text{(0)}}+\frac{\log H-\frac{i\pi}{2}}{2} \left(f_2^{{\rm ct},\text{(0)}} (\log H-\frac{i\pi}{2})+2 f_1^{{\rm ct},\text{(0)}}\right)\\
   &\qquad\qquad\, +\frac{n-2}{2} \left(f_1^{{\rm ct},\text{(0)}}+\left(\log H-\frac{i \pi}{2} \right)
   f_2^{{\rm ct},\text{(0)}}\right) \log \mu +\frac{(n-2)^2}{8}  f_2^{{\rm ct},\text{(0)}} \log
   ^2\mu ~.
\end{align}
\end{subequations}
Imposing the cancellation of divergences for arbitrary choices of the renormalization scale $\mu$, we arrive at $2\times(2+1)/2=3$ constraint equations,
\begin{subequations}
    \begin{align}
    0&=f_2^{{\rm ct},\text{(0)}}+f_2^{{\rm ct},\text{(1)}}+f_2^{\text{(2)}}~,\\
    0&=(n-2) f_2^{{\rm ct},\text{(0)}}+n f_2^{{\rm ct},\text{(1)}}+(n+2)
   f_2^{\text{(2)}}~,\\
    0&=\left(\log H -\frac{i\pi}{2}\right)
   \left(f_2^{{\rm ct},\text{(0)}}-f_2^{\text{(2)}}\right)+f_1^{{\rm ct},\text{(0)}}+f_1^{{\rm ct},\text{(1)}}
   +f_1^{\text{(2)}}~.
\end{align}
\end{subequations}
These allow us to solve the $2\times(2+1)/2=3$ coefficients $f_j^{{\rm ct},(l)}$ with $j\geq L-l$ as
\begin{subequations}
    \begin{align}
        f^{{\rm ct},(1)}_2&=-2f^{(2)}_2~,\\
        f^{{\rm ct},(1)}_1&=-f^{{\rm ct},(0)}_1-f^{(2)}_1~,\\
        f^{{\rm ct},(0)}_2&=f^{(2)}_2~.
    \end{align}
\end{subequations}
Inserting the solution back into the renormalized wavefunction coefficient yields
\begin{align}
    \nonumber\widehat\psi_n^{(2)}&=\psi_n^{(2)}+\psi_n^{{\rm ct},(1)}+\psi_n^{{\rm ct},(0)}\\
    &=f^{(2)}_0+f^{{\rm ct},(1)}_0+f^{{\rm ct},(0)}_0+\left(f^{(2)}_1-f^{{\rm ct},(0)}_1\right)\left(\log\frac{\mu}{H}+\frac{i\pi}{2}\right)+f^{(2)}_2\left(\log\frac{\mu}{H}+\frac{i\pi}{2}\right)^2~,
\end{align}
in agreement with \eqref{desiredForm}. Note that $\widehat\psi_n^{(2)}$ is not yet completely fixed due to the unspecified counterterm coefficients $f^{{\rm ct},(1)}_0$, $f^{{\rm ct},(0)}_0$ and $f^{{\rm ct},(0)}_1$. These counterterm coefficients are subjected to changes in the subtraction schemes and renormalization conditions. More generally, coefficient functions in the lower triangle, i.e. $f_j^{{\rm ct},(l)}$ with $j< L-l$, are degenerate with real shifts in $\log \mu$ and depend on schemes. By contrast, the coefficient functions in front of the leading logarithm, i.e. $f^{(L)}_j$, are diagram-specific and cannot be altered by schemes. 
We end this appendix by noting that this calculation is insensitive to specific Feynman diagrams and solely relies on their general structure and that the UV divergences can be removed by a finite number counterterms at a given order.

\allowdisplaybreaks
\section{Derivation of the correlator relations}\label{appendix3ptConsistencyRelation}

In this appendix, we present the detailed derivation of \eqref{2ptConsistencyRelation} and \eqref{3ptConsistencyRelation}. First, we recall our convention
\begin{equation}
    \log \Psi = \sum_{n=2}^{\infty}\frac{1}{n!}\left[\prod_{i=1}^n\int d^3k_i\ \varphi_{\vk_i}\right](2\pi)^3\delta(\vk_1+\ldots +\vk_n) \widehat\psi_n(\vk_1,\ldots,\vk_n)
\end{equation}
which implies, in the absence of parity violation such that  \smash{$\widehat\rho_n(\{\vk_i\}) \equiv \widehat\psi_n(\{\vk_i\})+\left[\widehat\psi_n(\{-\vk_i\})\right]^* =2\Re \widehat\psi_n(\{\vk_i\})$},
the path integral will be over the distribution
\begin{equation}
    \label{appB:realwfu}
    \log \abs{\Psi}^2 = \sum_{n=2}^{\infty}\frac{1}{n!}\left[\prod_{i=1}^n\int d^3k_i\ \varphi_{\vk_i}\right](2\pi)^3\delta(\vk_1+\ldots +\vk_n) \widehat\rho_n(\vk_1,\ldots,\vk_n).
\end{equation}
For the momentum correlators, we will also need
\begin{equation}
    \frac{\delta}{\delta\varphi_{-\vp}}\log\Psi=\sum_{n=2}^{\infty}\frac{1}{(n-1)!}\left[\prod_{i=1}^{n-1}\int d^3k_i\ \varphi_{\vk_i}\right](2\pi)^3\delta(\vk_1+\ldots +\vk_{n-1}-\vp) \widehat\psi_n(\vk_1,\ldots,\vk_{n-1},-\vp).
\end{equation}
To compute any correlator we would need to use the Born rule \eqref{BornRule} and divide by $\mathcal{N}\equiv \int D\varphi \abs{\Psi}^2$, but in focusing on the connected pieces we can eventually neglect these vacuum bubbles.
In particular, we restrict our attention to cubic theories with coupling $\lambda$ so that $n$-point, $L$-loop connected quantities $\widehat\psi_n$ scale as $\lambda^{n+2(L-1)}$. To derive 2- and 3-point functions up to 1-loop, we will thus only need the terms which go like
\begin{equation}
    \label{appB:WFUderivative}
    \begin{aligned}
     \frac{\delta}{\delta\varphi_{-\vp}}\log\Psi
     &=\left(\psi_2^{(0)}+\widehat\psi_2^{(1)}\right)\varphi_{-\vp}+\frac{1}{2}\int_{\vk_1\vk_2}\left(\psi_3^{(0)}+\widehat\psi_3^{(1)}\right)\varphi_{\vk_1}\varphi_{\vk_2}+\frac{1}{6}\int_{\vk_1\vk_2\vk_3}\psi_4^{(0)}\varphi_{\vk_1}\varphi_{\vk_2}\varphi_{\vk_3}\\
     &\qquad+\frac{1}{24}\int_{\vk_1\vk_2\vk_3\vk_4}\psi_5^{(0)}\varphi_{\vk_1}\varphi_{\vk_2}\varphi_{\vk_3}\varphi_{\vk_4}+\mathcal{O}(\lambda^4).
    \end{aligned}
\end{equation}
Let us start by calculating the 2-point relation \eqref{2ptConsistencyRelation} to one-loop order, so to order $\lambda^2$. We need the connected part of the correlators $\widehat B_2(p),\ \widehat C_2(p)$ defined in \eqref{eq:defCorrelators}.
For the 2-point correlator $\widehat B_2(p)$, the strategy is to expand the wavefunctional \eqref{appB:realwfu} around the Gaussian to order $\lambda^2$, keep only even powers of $\varphi$ in the path integral and perform the appropriate Wick contraction. 
Then we can use ad hoc Feynman rules for all the connected diagrams with free propagator $1/(-\rho_2^{(0)})$ and vertices $\widehat\rho_n$ to order $\lambda^2$ to calculate $\widehat B_2(p)$ in momentum space as
\begin{equation}
    \begin{aligned}
        \widehat B_2(p)&=
        \begin{tikzpicture}[baseline={([yshift=-.6ex]current bounding box.center)}, line cap=round]
            \useasboundingbox (0,-0.5) rectangle (1,0.5); 
            \draw[thick] (0,0) -- (1,0);
        \end{tikzpicture}
        +
        \begin{tikzpicture}[baseline={([yshift=-.6ex]current bounding box.center)}, line cap=round]
            \useasboundingbox (0,-0.5) rectangle (1,0.5);
            \draw[prop] (0,0) -- (1,0);
            \node[vertex] (M) at (0.5,0) {};
            \node[loopvertex] at (M) {};
        \end{tikzpicture}
        +
        \begin{tikzpicture}[baseline={([yshift=-.6ex]current bounding box.center)}, line cap=round]
            \useasboundingbox (0,-0.5) rectangle (1,0.5);
            \draw[prop] (0,0) -- (1,0);
            \node[vertex] (M) at (0.5,0) {};
            \draw[prop] (M) to[out=40, in=140, loop, looseness=30] (M);
        \end{tikzpicture}
        +
        \begin{tikzpicture}[baseline={([yshift=-.6ex]current bounding box.center)}, line cap=round]
            \useasboundingbox (0,-0.5) rectangle (1,0.5);
            \draw[prop] (0,0) -- (1,0);
            \node[vertex] (M1) at (0.5,0) {};
            \node[vertex] (M2) at (0.5,0.25) {};
            \draw[prop] (M1) -- (M2);
            \draw[prop] (M2) edge[out=40, in=140, loop, looseness=20] (M2);
        \end{tikzpicture}
        +
        \begin{tikzpicture}[baseline={([yshift=-.6ex]current bounding box.center)}, line cap=round]
            \useasboundingbox (0,-0.5) rectangle (1,0.5);
            \node[vertex] (M1) at (1/4, 0) {};
            \node[vertex] (M2) at (3/4, 0) {};
            \draw[prop] (0,0)  -- (M1);
            \draw[prop] (M2)   -- (1,0);
            \draw[prop] (M1) to[out=60,  in=120,  looseness=1.0] (M2);
            \draw[prop] (M1) to[out=-60, in=-120, looseness=1.0] (M2);
        \end{tikzpicture}
        +\mathcal{O}(\lambda^3)\\
        &=\frac{1}{-\rho_2^{(0)}(p)}+\frac{\widehat\rho_2^{(1)}(p)}{\left(-\rho_2^{(0)}(p)\right)^2}+\frac{1}{2}\int_{\vq}\frac{\rho_4^{(0)}(\vp,\vq,-\vq,-\vp)}{\left(-\rho_2^{(0)}(p)\right)^2\left(-\rho_2^{(0)}(q)\right)}\\
        &\ +\frac{1}{2}\frac{\rho_3^{(0)}(\vp,-\vp,0)}{\left(-\rho_2^{(0)}(p)\right)^2}\int_{\vq}\frac{\rho_3^{(0)}(\vq,-\vq,0)}{\left(-\rho_2^{(0)}(q)\right)\left(-\rho_2^{(0)}(0)\right)}+\frac{1}{2}\int_{\vq}\frac{\rho_3^{(0)}(\vp,\vq,-\vp-\vq)\rho_3^{(0)}(\vp+\vq,-\vq,-\vp)}{\left(-\rho_2^{(0)}(p)\right)^2\left(-\rho_2^{(0)}(q)\right)\left(-\rho_2^{(0)}(\abs{\vp+\vq})\right)}+\mathcal{O}(\lambda^3)
    \end{aligned}
\end{equation}
where we used solid lines to denote propagators, a dot to denote $\rho_n^{(0)}$ and a circle to differentiate the one-loop vertices $\widehat \rho_n^{(1)}$. Similarly, after taking the derivative \eqref{appB:WFUderivative} for the momentum correlator $\widehat C_2(p)$, we use dashed lines to represent the momentum leg and introduce a similar vertex notation corresponding to $\psi_n^{(0)},\ \widehat \psi_n^{(1)}$ so that 
\begin{equation}
    \begin{aligned}
        i \widehat C_2(p)&=\frac{1}{2}+
        \begin{tikzpicture}[baseline={([yshift=-.6ex]current bounding box.center)}, line cap=round]
            \useasboundingbox (0,-0.5) rectangle (1,0.5);
            \draw[prop] (0,0) -- (0.5,0);
            \draw[momentum] (0.5,0) -- (1.0,0);
            \node[vertex] (M) at (0.5,0) {};
            \node[loopvertex] at (M) {};
        \end{tikzpicture}
        +
        \begin{tikzpicture}[baseline={([yshift=-.6ex]current bounding box.center)}, line cap=round]
            \useasboundingbox (0,-0.5) rectangle (1,0.5);
            \draw[prop] (0,0) -- (0.5,0);
            \draw[momentum] (0.5,0) -- (1.0,0);
            \node[vertex] (M) at (0.5,0) {};
            \draw[prop] (M) to[out=40, in=140, loop, looseness=30] (M);
        \end{tikzpicture}
        +
        \begin{tikzpicture}[baseline={([yshift=-.6ex]current bounding box.center)}, line cap=round]
            \useasboundingbox (0,-0.5) rectangle (1,0.5);
            \draw[prop] (0,0) -- (0.5,0);
            \draw[momentum] (0.5,0) -- (1.0,0);
            \node[vertex] (M1) at (0.5,0) {};
            \node[vertex] (M2) at (0.5,0.25) {};
            \draw[prop] (M1) -- (M2);
            \draw[prop] (M2) edge[out=40, in=140, loop, looseness=20] (M2);
        \end{tikzpicture}
        +
        \begin{tikzpicture}[baseline={([yshift=-.6ex]current bounding box.center)}, line cap=round]
            \useasboundingbox (0,-0.5) rectangle (1,0.5);
            \node[vertex] (M1) at (1/4, 0) {};
            \node[vertex] (M2) at (3/4, 0) {};
            \draw[prop] (0,0)  -- (M1);
            \draw[momentum] (M2)   -- (1,0);
            \draw[prop] (M1) to[out=60,  in=120,  looseness=1.0] (M2);
            \draw[prop] (M1) to[out=-60, in=-120, looseness=1.0] (M2);
        \end{tikzpicture}
        +\psi_2^{(0)}(p)\widehat B_2(p)+\mathcal{O}(\lambda^3)\\
        &=\frac{1}{2}+\frac{\widehat\psi_2^{(1)}(p)}{\left(-\rho_2^{(0)}(p)\right)}+\frac{1}{2}\int_{\vq}\frac{\psi_4^{(0)}(\vp,\vq,-\vq,-\vp)}{\left(-\rho_2^{(0)}(p)\right)\left(-\rho_2^{(0)}(q)\right)}+\frac{1}{2}\frac{\psi_3^{(0)}(\vp,-\vp,0)}{-\rho_2^{(0)}(p)}\int_{\vq}\frac{\rho_3^{(0)}(\vq,-\vq,0)}{\left(-\rho_2^{(0)}(q)\right)\left(-\rho_2^{(0)}(p)\right)}\\
        &\qquad+\frac{1}{2}\int_{\vq}\frac{\rho_3^{(0)}(\vp,\vq,-\vp-\vq)\psi_3^{(0)}(\vp+\vq,-\vq,-\vp)}{\left(-\rho_2^{(0)}(p)\right)\left(-\rho_2^{(0)}(q)\right)\left(-\rho_2^{(0)}(\abs{\vp+\vq})\right)}+\psi_2^{(0)}(p)\widehat B_2(p)+\mathcal{O}(\lambda^3)\\
        &=\frac{1}{2}-\frac{\rho_2^{(0)}(p)}{2}\left(\widehat B_2(p)-\frac{1}{-\rho_2^{(0)}(p)}\right)+\psi_2^{(0)}(p)\widehat B_2
        (p)+\mathcal{O}(\lambda^3)=i \widehat \gamma_2^{(1)}(p)\widehat B_2(p)+\mathcal{O}(\lambda^3).
    \end{aligned}
\end{equation}
In the third equality, we used the fact that \smash{$\psi_n^{(0)}=\rho_n^{(0)}/2$} (because \smash{$\Im\widehat\psi_n=0$} at tree level) and then we related the diagrams to $\widehat B_2$ by noticing that they corresponded (up to an extra factor of $1/(-\rho_2^{(0)}(p))$) to all but the simple propagator term $B_2^{(0)}(p)$. The constant term $1/2$ sitting up front is a contact term from the anticommutator and is just conventional to define a convenient $\widehat C_2(p)$. 
One can now invert these relations to isolate $\widehat\rho_2^{(1)}(p)$ and $\widehat\gamma_2^{(1)}(p)$, and then act with $\tan(\frac{\pi}{2}\mu\del_\mu)\approx \frac{\pi}{2}\mu\del_\mu$ to find the relation between $\widehat B_2^{(1)},\widehat C_2^{(1)}$. In practice, it's simpler to act on $\widehat B_n^{(1)}$ with $\mu\del_\mu$, which will turn any $\widehat\rho_n^{(1)}$ into $\frac{4}{\pi}\widehat \gamma_n^{(1)}$, leaving us only to invert the equations for $\widehat\gamma_n^{(1)}$ the momentum correlators. (This is possible because classical loops are convergent.) Here we find
\begin{equation}
    \mu\del_\mu \widehat B_2(p)=\mu\del_\mu \widehat \rho_2^{(1)}(p)/\left(-\rho_2^{(0)}(p)\right)^2=\frac{4}{\pi}\widehat\gamma_2^{(1)}/\left(-\rho_2^{(0)}(p)\right)^2
\end{equation}
so that, noting $B_2^{(0)}(p)=1/(-\rho_2^{(0)}(p))$ and $\widehat C^{(1)}_2=(p)=\widehat \gamma_2^{(1)}(p)\widehat B_2^{(0)}(p)$, this reproduces \eqref{2ptConsistencyRelation}. We now perform the same steps for the 3-point correlators defined in \eqref{eq:defCorrelators}. First, we have
\begin{equation}
    \begin{aligned}
        \widehat B_3(\vp_1,\vp_2,\vp_3)&=
        \begin{tikzpicture}[baseline={([yshift=-.6ex]current bounding box.center)}, line cap=round]
            \useasboundingbox (0,-0.5) rectangle (1,0.5); 
            \coordinate (T1) at (0,0);
            \coordinate (T2) at (1.0,0.5);
            \coordinate (T3) at (1.0,-0.5);
            \coordinate (C) at (0.6,0);
            \draw[prop] (T1) -- (C);
            \draw[prop] (T2) -- (C);
            \draw[prop] (T3) -- (C);
            \node[vertex] (M) at (C){};
        \end{tikzpicture}
        +
        \begin{tikzpicture}[baseline={([yshift=-.6ex]current bounding box.center)}, line cap=round]
            \useasboundingbox (0,-0.5) rectangle (1,0.5); 
            \coordinate (T1) at (0,0);
            \coordinate (T2) at (1.0,0.5);
            \coordinate (T3) at (1.0,-0.5);
            \coordinate (C) at (0.6,0);
            \draw[prop] (T1) -- (C);
            \draw[prop] (T2) -- (C);
            \draw[prop] (T3) -- (C);
            \node[vertex] (M) at (C){};
            \node[loopvertex] (M) at (C){};
        \end{tikzpicture}
        +
        \begin{tikzpicture}[baseline={([yshift=-.6ex]current bounding box.center)}, line cap=round]
            \useasboundingbox (0,-0.5) rectangle (1,0.5); 
            \coordinate (T1) at (0,0);
            \coordinate (T2) at (1.0,0.5);
            \coordinate (T3) at (1.0,-0.5);

            \coordinate (T1half) at (0.3,0);
            \coordinate (T2half) at (0.8,0.25);
            \coordinate (T3half) at (0.8,-0.25);

            \draw[prop] (T1) -- (T1half);
            \draw[prop] (T2) -- (T2half);
            \draw[prop] (T3) -- (T3half);

            \draw[prop] (T1half) -- (T2half) -- (T3half) -- (T1half);
            
            \node[vertex] at (T1half){};
            \node[vertex] at (T2half){};
            \node[vertex] at (T3half){};
        \end{tikzpicture}
        +
        \begin{tikzpicture}[baseline={([yshift=-.6ex]current bounding box.center)}, line cap=round]
            \useasboundingbox (0,-0.5) rectangle (1,0.5); 
            \coordinate (T1) at (0,0);
            \coordinate (T2) at (1.0,0.5);
            \coordinate (T3) at (1.0,-0.5);
            \coordinate (C) at (0.6,0);
            \draw[prop] (T1) -- (C);
            \draw[prop] (T2) -- (C);
            \draw[prop] (T3) -- (C);
            \node[vertex] (M) at (C){};
            \node[vertex] (Mup) at (0.5,0.25){};
            \draw[prop] (C) -- (0.5,0.25);
            \draw[prop] (Mup) edge[out=40, in=140, loop, looseness=20] (Mup);
        \end{tikzpicture}
        +
        \begin{tikzpicture}[baseline={([yshift=-.6ex]current bounding box.center)}, line cap=round]
            \useasboundingbox (0,-0.5) rectangle (1,0.5); 
            \coordinate (T1) at (0,0);
            \coordinate (T2) at (1.0,0.5);
            \coordinate (T3) at (1.0,-0.5);
            \coordinate (C) at (0.6,0);
            \draw[prop] (T1) -- (C);
            \draw[prop] (T2) -- (C);
            \draw[prop] (T3) -- (C);
            \node[vertex] (M) at (C){};
            \draw[prop] (M) edge[out=80, in=170, loop, looseness=30] (M);
        \end{tikzpicture}
        \\
        &\qquad+\left[
        \begin{tikzpicture}[baseline={([yshift=-.6ex]current bounding box.center)}, line cap=round]
            \useasboundingbox (0,-0.5) rectangle (1,0.5); 
            \coordinate (T1) at (0,0);
            \coordinate (T2) at (1.0,0.5);
            \coordinate (T3) at (1.0,-0.5);
            \coordinate (C) at (0.6,0);
            \coordinate (rho2) at (0.3,0);
            \draw[prop] (T1) -- (C);
            \draw[prop] (T2) -- (C);
            \draw[prop] (T3) -- (C);
            \node[vertex] (M) at (C){};
            \node[vertex] at (rho2){};
            \node[loopvertex] at (rho2){};
        \end{tikzpicture}
        +
        \begin{tikzpicture}[baseline={([yshift=-.6ex]current bounding box.center)}, line cap=round]
            \useasboundingbox (0,-0.5) rectangle (1,0.5); 
            \coordinate (T1) at (0,0);
            \coordinate (T2) at (1.0,0.5);
            \coordinate (T3) at (1.0,-0.5);
            \coordinate (C) at (0.6,0);
            \coordinate (rho3) at (0.3,0);
            \coordinate (rho3above) at (0.3,0.25);
            \draw[prop] (T1) -- (C);
            \draw[prop] (T2) -- (C);
            \draw[prop] (T3) -- (C);
            \node[vertex] (M) at (C){};
            \node[vertex] at (rho3){};
            \node[vertex] (M2) at (rho3above){};
            \draw[prop] (rho3) -- (rho3above);
            \draw[prop] (M2) edge[out=40, in=140, loop, looseness=20] (M2);
        \end{tikzpicture}
        +
        \begin{tikzpicture}[baseline={([yshift=-.6ex]current bounding box.center)}, line cap=round]
            \useasboundingbox (0,-0.5) rectangle (1,0.5); 
            \coordinate (T1) at (0,0);
            \coordinate (T2) at (1.0,0.5);
            \coordinate (T3) at (1.0,-0.5);
            \coordinate (T1a) at (0.15,0);
            \coordinate (T1b) at (0.45,0);
            
            \coordinate (C) at (0.6,0);
            \draw[prop] (T1) -- (T1a);
            \draw[prop] (T1b) -- (C);
            \draw[prop] (T2) -- (C);
            \draw[prop] (T3) -- (C);
            \node[vertex] (M) at (C){};
            \node[vertex] (M1) at (T1a){};
            \node[vertex] (M2) at (T1b){};

            \draw[prop] (M1) to[out=60,  in=120,  looseness=1.0] (M2);
            \draw[prop] (M1) to[out=-60, in=-120, looseness=1.0] (M2);
        \end{tikzpicture}
        +
        \begin{tikzpicture}[baseline={([yshift=-.6ex]current bounding box.center)}, line cap=round]
            \useasboundingbox (0,-0.5) rectangle (1,0.5); 
            \coordinate (T1) at (0,0);
            \coordinate (T2) at (1.0,0.5);
            \coordinate (T3) at (1.0,-0.5);
            \coordinate (C) at (0.6,0);
            \coordinate (rho3) at (0.3,0);
            \draw[prop] (T1) -- (C);
            \draw[prop] (T2) -- (C);
            \draw[prop] (T3) -- (C);
            \node[vertex] (M) at (C){};
            \node[vertex] (M2) at (rho3){};
            \draw[prop] (M2) edge[out=40, in=140, loop, looseness=20] (M2);
        \end{tikzpicture}
        +
        \begin{tikzpicture}[baseline={([yshift=-.6ex]current bounding box.center)}, line cap=round]
            \useasboundingbox (0,-0.5) rectangle (1,0.5); 
            \coordinate (T1) at (0,0);
            \coordinate (T2) at (1.0,0.5);
            \coordinate (T3) at (1.0,-0.5);
            \coordinate (T1a) at (0.15,0);
            
            \coordinate (C) at (0.6,0);
            \draw[prop] (T1) -- (T1a);
            \draw[prop] (T2) -- (C);
            \draw[prop] (T3) -- (C);
            \node[vertex] (M) at (C){};
            \node[vertex] (M1) at (T1a){};

            \draw[prop] (M1) to[out=60,  in=120,  looseness=1.0] (M);
            \draw[prop] (M1) to[out=-60, in=-120, looseness=1.0] (M);
        \end{tikzpicture}
        +(1\leftrightarrow2)+(1\leftrightarrow3)\right]+\mathcal{O}(\lambda^4)
    \end{aligned}
\end{equation}
while
\begin{equation}
    \begin{aligned}
       i \widehat C_3(\vp_1,\vp_2;\vp_3)&=
        \begin{tikzpicture}[baseline={([yshift=-.6ex]current bounding box.center)}, line cap=round]
            \useasboundingbox (0,-0.5) rectangle (1,0.5); 
            \coordinate (T1) at (0,0);
            \coordinate (T2) at (1.0,0.5);
            \coordinate (T3) at (1.0,-0.5);
            \coordinate (C) at (0.6,0);
            \draw[prop] (T1) -- (C);
            \draw[prop] (T2) -- (C);
            \draw[momentum] (T3) -- (C);
            \node[vertex] (M) at (C){};
        \end{tikzpicture}
        +
        \begin{tikzpicture}[baseline={([yshift=-.6ex]current bounding box.center)}, line cap=round]
            \useasboundingbox (0,-0.5) rectangle (1,0.5); 
            \coordinate (T1) at (0,0);
            \coordinate (T2) at (1.0,0.5);
            \coordinate (T3) at (1.0,-0.5);
            \coordinate (C) at (0.6,0);
            \draw[prop] (T1) -- (C);
            \draw[prop] (T2) -- (C);
            \draw[momentum] (T3) -- (C);
            \node[vertex] (M) at (C){};
            \node[loopvertex] (M) at (C){};
        \end{tikzpicture}
        +
        \begin{tikzpicture}[baseline={([yshift=-.6ex]current bounding box.center)}, line cap=round]
            \useasboundingbox (0,-0.5) rectangle (1,0.5); 
            \coordinate (T1) at (0,0);
            \coordinate (T2) at (1.0,0.5);
            \coordinate (T3) at (1.0,-0.5);

            \coordinate (T1half) at (0.3,0);
            \coordinate (T2half) at (0.8,0.25);
            \coordinate (T3half) at (0.8,-0.25);

            \draw[prop] (T1) -- (T1half);
            \draw[prop] (T2) -- (T2half);
            \draw[momentum] (T3) -- (T3half);

            \draw[prop] (T1half) -- (T2half) -- (T3half) -- (T1half);
            
            \node[vertex] at (T1half){};
            \node[vertex] at (T2half){};
            \node[vertex] at (T3half){};
        \end{tikzpicture}
        +
        \begin{tikzpicture}[baseline={([yshift=-.6ex]current bounding box.center)}, line cap=round]
            \useasboundingbox (0,-0.5) rectangle (1,0.5); 
            \coordinate (T1) at (0,0);
            \coordinate (T2) at (1.0,0.5);
            \coordinate (T3) at (1.0,-0.5);
            \coordinate (C) at (0.6,0);
            \draw[prop] (T1) -- (C);
            \draw[prop] (T2) -- (C);
            \draw[momentum] (T3) -- (C);
            \node[vertex] (M) at (C){};
            \node[vertex] (Mup) at (0.5,0.25){};
            \draw[prop] (C) -- (0.5,0.25);
            \draw[prop] (Mup) edge[out=40, in=140, loop, looseness=20] (Mup);
        \end{tikzpicture}
        +
        \begin{tikzpicture}[baseline={([yshift=-.6ex]current bounding box.center)}, line cap=round]
            \useasboundingbox (0,-0.5) rectangle (1,0.5); 
            \coordinate (T1) at (0,0);
            \coordinate (T2) at (1.0,0.5);
            \coordinate (T3) at (1.0,-0.5);
            \coordinate (C) at (0.6,0);
            \draw[prop] (T1) -- (C);
            \draw[prop] (T2) -- (C);
            \draw[momentum] (T3) -- (C);
            \node[vertex] (M) at (C){};
            \draw[prop] (M) edge[out=80, in=170, loop, looseness=30] (M);
        \end{tikzpicture}
        +\left[
        \begin{tikzpicture}[baseline={([yshift=-.6ex]current bounding box.center)}, line cap=round]
            \useasboundingbox (0,-0.5) rectangle (1,0.5); 
            \coordinate (T1) at (0,0);
            \coordinate (T2) at (1.0,0.5);
            \coordinate (T3) at (1.0,-0.5);
            \coordinate (C) at (0.6,0);
            \coordinate (rho2) at (0.8,-0.25);
            \draw[prop] (T1) -- (C);
            \draw[prop] (T2) -- (C);
            \draw[momentum] (T3) -- (rho2);
            \draw[prop] (rho2) -- (C);
            \node[vertex] (M) at (C){};
            \node[vertex] at (rho2){};
            \node[loopvertex] at (rho2){};
        \end{tikzpicture}
        +
        \begin{tikzpicture}[baseline={([yshift=-.6ex]current bounding box.center)}, line cap=round]
            \useasboundingbox (0,-0.5) rectangle (1,0.5); 
            \coordinate (T1) at (0,0);
            \coordinate (T2) at (1.0,0.5);
            \coordinate (T3) at (1.0,-0.5);
            \coordinate (C) at (0.6,0);
            \coordinate (rho3) at (0.8,-0.25);
            \coordinate (rho3above) at (0.4,-0.25);
            \draw[prop] (T1) -- (C);
            \draw[prop] (T2) -- (C);
            \draw[prop] (rho3) -- (C);
            \draw[momentum] (T3) -- (rho3);
            \node[vertex] (M) at (C){};
            \node[vertex] at (rho3){};
            \node[vertex] (M2) at (rho3above){};
            \draw[prop] (rho3) -- (rho3above);
            \draw[prop] (M2) edge[out=130, in=230, loop, looseness=20] (M2);
        \end{tikzpicture}
        +
        \begin{tikzpicture}[baseline={([yshift=-.6ex]current bounding box.center)}, line cap=round]
            \useasboundingbox (0,-0.5) rectangle (1,0.5); 
            \coordinate (T1) at (0,0);
            \coordinate (T2) at (1.0,0.5);
            \coordinate (T3) at (1.0,-0.5);
            \coordinate (T3a) at (0.9,-0.375);
            \coordinate (T3b) at (0.72,-0.15);
            
            \coordinate (C) at (0.6,0);
            \draw[momentum] (T3) -- (T3a);
            \draw[prop] (T3b) -- (C);
            \draw[prop] (T2) -- (C);
            \draw[prop] (T1) -- (C);
            \node[vertex] (M) at (C){};
            \node[vertex] (M1) at (T3a){};
            \node[vertex] (M2) at (T3b){};

            \draw[prop] (M1) to[out=60,  in=0,  looseness=1.0] (M2);
            \draw[prop] (M1) to[out=-180, in=-120, looseness=1.0] (M2);
        \end{tikzpicture}
        +
        \begin{tikzpicture}[baseline={([yshift=-.6ex]current bounding box.center)}, line cap=round]
            \useasboundingbox (0,-0.5) rectangle (1,0.5); 
            \coordinate (T1) at (0,0);
            \coordinate (T2) at (1.0,0.5);
            \coordinate (T3) at (1.0,-0.5);
            \coordinate (C) at (0.6,0);
            \coordinate (rho3) at (0.8,-0.25);
            \draw[prop] (T1) -- (C);
            \draw[prop] (T2) -- (C);
            \draw[prop] (rho3) -- (C);
            \draw[momentum] (T3) -- (rho3);
            \node[vertex] (M) at (C){};
            \node[vertex] (M2) at (rho3){};
            \draw[prop] (M2) edge[out=160, in=260, loop, looseness=20] (M2);
        \end{tikzpicture}
        +
        \begin{tikzpicture}[baseline={([yshift=-.6ex]current bounding box.center)}, line cap=round]
            \useasboundingbox (0,-0.5) rectangle (1,0.5); 
            \coordinate (T1) at (0,0);
            \coordinate (T2) at (1.0,0.5);
            \coordinate (T3) at (1.0,-0.5);
            \coordinate (rho3) at (0.9,-0.375);
            
            \coordinate (C) at (0.6,0);
            \draw[momentum] (T3) -- (rho3);
            \draw[prop] (T1) -- (C);
            \draw[prop] (T2) -- (C);
            \node[vertex] (M) at (C){};
            \node[vertex] (M1) at (rho3){};

            \draw[prop] (M1) to[out=60,  in=0,  looseness=1.0] (M);
            \draw[prop] (M1) to[out=-180, in=-120, looseness=1.0] (M);
        \end{tikzpicture}\right]
        \\
        &\qquad+\left[
        \begin{tikzpicture}[baseline={([yshift=-.6ex]current bounding box.center)}, line cap=round]
            \useasboundingbox (0,-0.5) rectangle (1,0.5); 
            \coordinate (T1) at (0,0);
            \coordinate (T2) at (1.0,0.5);
            \coordinate (T3) at (1.0,-0.5);
            \coordinate (C) at (0.6,0);
            \coordinate (rho2) at (0.3,0);
            \draw[prop] (T1) -- (C);
            \draw[prop] (T2) -- (C);
            \draw[momentum] (T3) -- (C);
            \node[vertex] (M) at (C){};
            \node[vertex] at (rho2){};
            \node[loopvertex] at (rho2){};
        \end{tikzpicture}
        +
        \begin{tikzpicture}[baseline={([yshift=-.6ex]current bounding box.center)}, line cap=round]
            \useasboundingbox (0,-0.5) rectangle (1,0.5); 
            \coordinate (T1) at (0,0);
            \coordinate (T2) at (1.0,0.5);
            \coordinate (T3) at (1.0,-0.5);
            \coordinate (C) at (0.6,0);
            \coordinate (rho3) at (0.3,0);
            \coordinate (rho3above) at (0.3,0.25);
            \draw[prop] (T1) -- (C);
            \draw[prop] (T2) -- (C);
            \draw[momentum] (T3) -- (C);
            \node[vertex] (M) at (C){};
            \node[vertex] at (rho3){};
            \node[vertex] (M2) at (rho3above){};
            \draw[prop] (rho3) -- (rho3above);
            \draw[prop] (M2) edge[out=40, in=140, loop, looseness=20] (M2);
        \end{tikzpicture}
        +
        \begin{tikzpicture}[baseline={([yshift=-.6ex]current bounding box.center)}, line cap=round]
            \useasboundingbox (0,-0.5) rectangle (1,0.5); 
            \coordinate (T1) at (0,0);
            \coordinate (T2) at (1.0,0.5);
            \coordinate (T3) at (1.0,-0.5);
            \coordinate (T1a) at (0.15,0);
            \coordinate (T1b) at (0.45,0);
            
            \coordinate (C) at (0.6,0);
            \draw[prop] (T1) -- (T1a);
            \draw[prop] (T1b) -- (C);
            \draw[prop] (T2) -- (C);
            \draw[momentum] (T3) -- (C);
            \node[vertex] (M) at (C){};
            \node[vertex] (M1) at (T1a){};
            \node[vertex] (M2) at (T1b){};

            \draw[prop] (M1) to[out=60,  in=120,  looseness=1.0] (M2);
            \draw[prop] (M1) to[out=-60, in=-120, looseness=1.0] (M2);
        \end{tikzpicture}
        +
        \begin{tikzpicture}[baseline={([yshift=-.6ex]current bounding box.center)}, line cap=round]
            \useasboundingbox (0,-0.5) rectangle (1,0.5); 
            \coordinate (T1) at (0,0);
            \coordinate (T2) at (1.0,0.5);
            \coordinate (T3) at (1.0,-0.5);
            \coordinate (C) at (0.6,0);
            \coordinate (rho3) at (0.3,0);
            \draw[prop] (T1) -- (C);
            \draw[prop] (T2) -- (C);
            \draw[momentum] (T3) -- (C);
            \node[vertex] (M) at (C){};
            \node[vertex] (M2) at (rho3){};
            \draw[prop] (M2) edge[out=40, in=140, loop, looseness=20] (M2);
        \end{tikzpicture}
        +
        \begin{tikzpicture}[baseline={([yshift=-.6ex]current bounding box.center)}, line cap=round]
            \useasboundingbox (0,-0.5) rectangle (1,0.5); 
            \coordinate (T1) at (0,0);
            \coordinate (T2) at (1.0,0.5);
            \coordinate (T3) at (1.0,-0.5);
            \coordinate (T1a) at (0.15,0);
            
            \coordinate (C) at (0.6,0);
            \draw[prop] (T1) -- (T1a);
            \draw[prop] (T2) -- (C);
            \draw[momentum] (T3) -- (C);
            \node[vertex] (M) at (C){};
            \node[vertex] (M1) at (T1a){};

            \draw[prop] (M1) to[out=60,  in=120,  looseness=1.0] (M);
            \draw[prop] (M1) to[out=-60, in=-120, looseness=1.0] (M);
        \end{tikzpicture}
        +(1\leftrightarrow2)\right]+\psi_2^{(0)}(p_3)\widehat B_3(\vp_1,\vp_2,\vp_3) +\mathcal{O}(\lambda^4)\\
        &=-\frac{\rho_2^{(0)}(p_3)}{2}\widehat B_3(\vp_1,\vp_2,\vp_3)+\frac{i \widehat\gamma^{(1)}_3(\vp_1,\vp_2,\vp_3)}{\left(-\rho_2^{(0)}(p_1)\right)\left(-\rho_2^{(0)}(p_2)\right)}+\psi_2^{(0)}(p_3)\widehat B_3(\vp_1,\vp_2,\vp_3) +\mathcal{O}(\lambda^4)\\
        &=i\widehat\gamma^{(1)}_2(p_3)B_3^{(0)}(\vp_1,\vp_2,\vp_3)+i\widehat\gamma^{(1)}_3(\vp_1,\vp_2,\vp_3)B_2^{(0)}(p_1)B_2^{(0)}(p_2)+\mathcal{O}(\lambda^4).
    \end{aligned}
\end{equation}
In the second equality we again used that all diagrams in $\widehat C_3$ have a partner in $\widehat B_3$ up to a factor of $(-\rho_2^{(0)}(p_3))/2$. To get to the last equality, we used the explicit expressions for $B_2^{(0)},\ B_3^{(0)}$. Finally, we solve for $\widehat\gamma_3^{(1)}$ and retrieve the equation \eqref{3ptConsistencyRelation} using
\begin{equation}
    \begin{aligned}
        \mu\del_\mu \widehat B_3^{(1)}(\vp_1,\vp_2,\vp_3)&=\mu\del_\mu\left[
        \begin{tikzpicture}[baseline={([yshift=-.6ex]current bounding box.center)}, line cap=round]
            \useasboundingbox (0,-0.5) rectangle (1,0.5); 
            \coordinate (T1) at (0,0);
            \coordinate (T2) at (1.0,0.5);
            \coordinate (T3) at (1.0,-0.5);
            \coordinate (C) at (0.6,0);
            \draw[prop] (T1) -- (C);
            \draw[prop] (T2) -- (C);
            \draw[prop] (T3) -- (C);
            \node[vertex] (M) at (C){};
            \node[loopvertex] (M) at (C){};
        \end{tikzpicture}+\begin{tikzpicture}[baseline={([yshift=-.6ex]current bounding box.center)}, line cap=round]
            \useasboundingbox (0,-0.5) rectangle (1,0.5); 
            \coordinate (T1) at (0,0);
            \coordinate (T2) at (1.0,0.5);
            \coordinate (T3) at (1.0,-0.5);
            \coordinate (C) at (0.6,0);
            \coordinate (rho2) at (0.3,0);
            \draw[prop] (T1) -- (C);
            \draw[prop] (T2) -- (C);
            \draw[prop] (T3) -- (C);
            \node[vertex] (M) at (C){};
            \node[vertex] at (rho2){};
            \node[loopvertex] at (rho2){};
        \end{tikzpicture}
        +
        \begin{tikzpicture}[baseline={([yshift=-.6ex]current bounding box.center)}, line cap=round]
            \useasboundingbox (0,-0.5) rectangle (1,0.5); 
            \coordinate (T1) at (0,0);
            \coordinate (T2) at (1.0,0.5);
            \coordinate (T3) at (1.0,-0.5);
            \coordinate (C) at (0.6,0);
            \coordinate (rho2) at (0.8,+0.25);
            \draw[prop] (T1) -- (C);
            \draw[prop] (T2) -- (C);
            \draw[prop] (T3) -- (C);
            \node[vertex] (M) at (C){};
            \node[vertex] at (rho2){};
            \node[loopvertex] at (rho2){};
        \end{tikzpicture}
        +
        \begin{tikzpicture}[baseline={([yshift=-.6ex]current bounding box.center)}, line cap=round]
            \useasboundingbox (0,-0.5) rectangle (1,0.5); 
            \coordinate (T1) at (0,0);
            \coordinate (T2) at (1.0,0.5);
            \coordinate (T3) at (1.0,-0.5);
            \coordinate (C) at (0.6,0);
            \coordinate (rho2) at (0.8,-0.25);
            \draw[prop] (T1) -- (C);
            \draw[prop] (T2) -- (C);
            \draw[prop] (T3) -- (C);
            \node[vertex] (M) at (C){};
            \node[vertex] at (rho2){};
            \node[loopvertex] at (rho2){};
        \end{tikzpicture}\right]\\
        &=\frac{4}{\pi}\frac{\left[\widehat\gamma_3^{(1)}(\vp_1,\vp_2,\vp_3)+\sum_{i=1}^3\widehat\gamma_2^{(1)}(p_i)/\left(-\rho_2^{(0)}(p_i)\right)\right]}{\left(-\rho_2^{(0)}(p_1)\right)\left(-\rho_2^{(0)}(p_2)\right)\left(-\rho_2^{(0)}(p_3)\right)}\\
        &=\frac{4}{\pi}\left[B_2^{(0)}(p_3)\widehat  C_3^{(1)}(\vp_1,\vp_2;\vp_3)+B_3^{(0)}(\vp_1,\vp_2,\vp_3)\left(\widehat C_2^{(1)}(p_1)+\widehat C_2^{(1)}(p_2)\right)\right].
    \end{aligned}
\end{equation}

\twocolumngrid


\bibliographystyle{apsrev4-2}

\bibliography{Refs}

\begin{thebibliography}{82}%
\makeatletter
\providecommand \@ifxundefined [1]{%
 \@ifx{#1\undefined}
}%
\providecommand \@ifnum [1]{%
 \ifnum #1\expandafter \@firstoftwo
 \else \expandafter \@secondoftwo
 \fi
}%
\providecommand \@ifx [1]{%
 \ifx #1\expandafter \@firstoftwo
 \else \expandafter \@secondoftwo
 \fi
}%
\providecommand \natexlab [1]{#1}%
\providecommand \enquote  [1]{``#1''}%
\providecommand \bibnamefont  [1]{#1}%
\providecommand \bibfnamefont [1]{#1}%
\providecommand \citenamefont [1]{#1}%
\providecommand \href@noop [0]{\@secondoftwo}%
\providecommand \href [0]{\begingroup \@sanitize@url \@href}%
\providecommand \@href[1]{\@@startlink{#1}\@@href}%
\providecommand \@@href[1]{\endgroup#1\@@endlink}%
\providecommand \@sanitize@url [0]{\catcode `\\12\catcode `\$12\catcode
  `\&12\catcode `\#12\catcode `\^12\catcode `\_12\catcode `\%12\relax}%
\providecommand \@@startlink[1]{}%
\providecommand \@@endlink[0]{}%
\providecommand \url  [0]{\begingroup\@sanitize@url \@url }%
\providecommand \@url [1]{\endgroup\@href {#1}{\urlprefix }}%
\providecommand \urlprefix  [0]{URL }%
\providecommand \Eprint [0]{\href }%
\providecommand \doibase [0]{https://doi.org/}%
\providecommand \selectlanguage [0]{\@gobble}%
\providecommand \bibinfo  [0]{\@secondoftwo}%
\providecommand \bibfield  [0]{\@secondoftwo}%
\providecommand \translation [1]{[#1]}%
\providecommand \BibitemOpen [0]{}%
\providecommand \bibitemStop [0]{}%
\providecommand \bibitemNoStop [0]{.\EOS\space}%
\providecommand \EOS [0]{\spacefactor3000\relax}%
\providecommand \BibitemShut  [1]{\csname bibitem#1\endcsname}%
\let\auto@bib@innerbib\@empty
\bibitem [{\citenamefont {Baume}\ \emph {et~al.}(2014)\citenamefont {Baume},
  \citenamefont {Keren-Zur}, \citenamefont {Rattazzi},\ and\ \citenamefont
  {Vitale}}]{localCS}%
  \BibitemOpen
  \bibfield  {author} {\bibinfo {author} {\bibfnamefont {F.}~\bibnamefont
  {Baume}}, \bibinfo {author} {\bibfnamefont {B.}~\bibnamefont {Keren-Zur}},
  \bibinfo {author} {\bibfnamefont {R.}~\bibnamefont {Rattazzi}},\ and\
  \bibinfo {author} {\bibfnamefont {L.}~\bibnamefont {Vitale}},\ }\bibfield
  {journal} {\bibinfo  {journal} {Journal of High Energy Physics}\ }\textbf
  {\bibinfo {volume} {2014}},\ \href {https://doi.org/10.1007/jhep08(2014)152}
  {10.1007/jhep08(2014)152} (\bibinfo {year} {2014})\BibitemShut {NoStop}%
\bibitem [{\citenamefont {Liu}\ \emph {et~al.}(2020)\citenamefont {Liu},
  \citenamefont {Tong}, \citenamefont {Wang},\ and\ \citenamefont
  {Xianyu}}]{Liu:2019fag}%
  \BibitemOpen
  \bibfield  {author} {\bibinfo {author} {\bibfnamefont {T.}~\bibnamefont
  {Liu}}, \bibinfo {author} {\bibfnamefont {X.}~\bibnamefont {Tong}}, \bibinfo
  {author} {\bibfnamefont {Y.}~\bibnamefont {Wang}},\ and\ \bibinfo {author}
  {\bibfnamefont {Z.-Z.}\ \bibnamefont {Xianyu}},\ }\href
  {https://doi.org/10.1007/JHEP04(2020)189} {\bibfield  {journal} {\bibinfo
  {journal} {JHEP}\ }\textbf {\bibinfo {volume} {04}},\ \bibinfo {pages}
  {189}},\ \Eprint {https://arxiv.org/abs/1909.01819} {arXiv:1909.01819
  [hep-ph]} \BibitemShut {NoStop}%
\bibitem [{\citenamefont {Cabass}\ \emph {et~al.}(2023)\citenamefont {Cabass},
  \citenamefont {Jazayeri}, \citenamefont {Pajer},\ and\ \citenamefont
  {Stefanyszyn}}]{Cabass:2022rhr}%
  \BibitemOpen
  \bibfield  {author} {\bibinfo {author} {\bibfnamefont {G.}~\bibnamefont
  {Cabass}}, \bibinfo {author} {\bibfnamefont {S.}~\bibnamefont {Jazayeri}},
  \bibinfo {author} {\bibfnamefont {E.}~\bibnamefont {Pajer}},\ and\ \bibinfo
  {author} {\bibfnamefont {D.}~\bibnamefont {Stefanyszyn}},\ }\href
  {https://doi.org/10.1007/JHEP02(2023)021} {\bibfield  {journal} {\bibinfo
  {journal} {JHEP}\ }\textbf {\bibinfo {volume} {02}},\ \bibinfo {pages}
  {021}},\ \Eprint {https://arxiv.org/abs/2210.02907} {arXiv:2210.02907
  [hep-th]} \BibitemShut {NoStop}%
\bibitem [{\citenamefont {Goodhew}\ \emph {et~al.}(2024)\citenamefont
  {Goodhew}, \citenamefont {Thavanesan},\ and\ \citenamefont
  {Wall}}]{Goodhew:2024eup}%
  \BibitemOpen
  \bibfield  {author} {\bibinfo {author} {\bibfnamefont {H.}~\bibnamefont
  {Goodhew}}, \bibinfo {author} {\bibfnamefont {A.}~\bibnamefont
  {Thavanesan}},\ and\ \bibinfo {author} {\bibfnamefont {A.~C.}\ \bibnamefont
  {Wall}},\ }\href@noop {} {\  (\bibinfo {year} {2024})},\ \Eprint
  {https://arxiv.org/abs/2408.17406} {arXiv:2408.17406 [hep-th]} \BibitemShut
  {NoStop}%
\bibitem [{\citenamefont {Thavanesan}(2025)}]{Thavanesan:2025kyc}%
  \BibitemOpen
  \bibfield  {author} {\bibinfo {author} {\bibfnamefont {A.}~\bibnamefont
  {Thavanesan}},\ }\href@noop {} {\  (\bibinfo {year} {2025})},\ \Eprint
  {https://arxiv.org/abs/2501.06383} {arXiv:2501.06383 [hep-th]} \BibitemShut
  {NoStop}%
\bibitem [{\citenamefont {Lee}\ \emph {et~al.}(2023)\citenamefont {Lee},
  \citenamefont {McCulloch},\ and\ \citenamefont {Pajer}}]{Lee:2023jby}%
  \BibitemOpen
  \bibfield  {author} {\bibinfo {author} {\bibfnamefont {M.~H.~G.}\
  \bibnamefont {Lee}}, \bibinfo {author} {\bibfnamefont {C.}~\bibnamefont
  {McCulloch}},\ and\ \bibinfo {author} {\bibfnamefont {E.}~\bibnamefont
  {Pajer}},\ }\href {https://doi.org/10.1007/JHEP11(2023)038} {\bibfield
  {journal} {\bibinfo  {journal} {JHEP}\ }\textbf {\bibinfo {volume} {11}},\
  \bibinfo {pages} {038}},\ \Eprint {https://arxiv.org/abs/2305.11228}
  {arXiv:2305.11228 [hep-th]} \BibitemShut {NoStop}%
\bibitem [{\citenamefont {Jain}\ \emph {et~al.}(2025)\citenamefont {Jain},
  \citenamefont {Pajer},\ and\ \citenamefont {Tong}}]{Jain:2025maa}%
  \BibitemOpen
  \bibfield  {author} {\bibinfo {author} {\bibfnamefont {D.}~\bibnamefont
  {Jain}}, \bibinfo {author} {\bibfnamefont {E.}~\bibnamefont {Pajer}},\ and\
  \bibinfo {author} {\bibfnamefont {X.}~\bibnamefont {Tong}},\ }\href@noop {}
  {\  (\bibinfo {year} {2025})},\ \Eprint {https://arxiv.org/abs/2509.02696}
  {arXiv:2509.02696 [hep-th]} \BibitemShut {NoStop}%
\bibitem [{\citenamefont {Tsamis}\ and\ \citenamefont
  {Woodard}(1996)}]{Tsamis:1996qq}%
  \BibitemOpen
  \bibfield  {author} {\bibinfo {author} {\bibfnamefont {N.~C.}\ \bibnamefont
  {Tsamis}}\ and\ \bibinfo {author} {\bibfnamefont {R.~P.}\ \bibnamefont
  {Woodard}},\ }\href {https://doi.org/10.1016/0550-3213(96)00246-5} {\bibfield
   {journal} {\bibinfo  {journal} {Nucl. Phys. B}\ }\textbf {\bibinfo {volume}
  {474}},\ \bibinfo {pages} {235} (\bibinfo {year} {1996})},\ \Eprint
  {https://arxiv.org/abs/hep-ph/9602315} {arXiv:hep-ph/9602315} \BibitemShut
  {NoStop}%
\bibitem [{\citenamefont {Weinberg}(2005)}]{Weinberg:2005vy}%
  \BibitemOpen
  \bibfield  {author} {\bibinfo {author} {\bibfnamefont {S.}~\bibnamefont
  {Weinberg}},\ }\href {https://doi.org/10.1103/PhysRevD.72.043514} {\bibfield
  {journal} {\bibinfo  {journal} {Phys. Rev. D}\ }\textbf {\bibinfo {volume}
  {72}},\ \bibinfo {pages} {043514} (\bibinfo {year} {2005})},\ \Eprint
  {https://arxiv.org/abs/hep-th/0506236} {arXiv:hep-th/0506236} \BibitemShut
  {NoStop}%
\bibitem [{\citenamefont {Marolf}\ and\ \citenamefont
  {Morrison}(2010)}]{Marolf:2010zp}%
  \BibitemOpen
  \bibfield  {author} {\bibinfo {author} {\bibfnamefont {D.}~\bibnamefont
  {Marolf}}\ and\ \bibinfo {author} {\bibfnamefont {I.~A.}\ \bibnamefont
  {Morrison}},\ }\href {https://doi.org/10.1103/PhysRevD.82.105032} {\bibfield
  {journal} {\bibinfo  {journal} {Phys. Rev. D}\ }\textbf {\bibinfo {volume}
  {82}},\ \bibinfo {pages} {105032} (\bibinfo {year} {2010})},\ \Eprint
  {https://arxiv.org/abs/1006.0035} {arXiv:1006.0035 [gr-qc]} \BibitemShut
  {NoStop}%
\bibitem [{\citenamefont {Krotov}\ and\ \citenamefont
  {Polyakov}(2011)}]{Krotov:2010ma}%
  \BibitemOpen
  \bibfield  {author} {\bibinfo {author} {\bibfnamefont {D.}~\bibnamefont
  {Krotov}}\ and\ \bibinfo {author} {\bibfnamefont {A.~M.}\ \bibnamefont
  {Polyakov}},\ }\href {https://doi.org/10.1016/j.nuclphysb.2011.03.025}
  {\bibfield  {journal} {\bibinfo  {journal} {Nucl. Phys. B}\ }\textbf
  {\bibinfo {volume} {849}},\ \bibinfo {pages} {410} (\bibinfo {year}
  {2011})},\ \Eprint {https://arxiv.org/abs/1012.2107} {arXiv:1012.2107
  [hep-th]} \BibitemShut {NoStop}%
\bibitem [{\citenamefont {Bros}\ \emph {et~al.}(2008)\citenamefont {Bros},
  \citenamefont {Epstein},\ and\ \citenamefont {Moschella}}]{Bros:2006gs}%
  \BibitemOpen
  \bibfield  {author} {\bibinfo {author} {\bibfnamefont {J.}~\bibnamefont
  {Bros}}, \bibinfo {author} {\bibfnamefont {H.}~\bibnamefont {Epstein}},\ and\
  \bibinfo {author} {\bibfnamefont {U.}~\bibnamefont {Moschella}},\ }\href
  {https://doi.org/10.1088/1475-7516/2008/02/003} {\bibfield  {journal}
  {\bibinfo  {journal} {JCAP}\ }\textbf {\bibinfo {volume} {02}},\ \bibinfo
  {pages} {003}},\ \Eprint {https://arxiv.org/abs/hep-th/0612184}
  {arXiv:hep-th/0612184} \BibitemShut {NoStop}%
\bibitem [{\citenamefont {Senatore}\ and\ \citenamefont
  {Zaldarriaga}(2010)}]{Senatore:2009cf}%
  \BibitemOpen
  \bibfield  {author} {\bibinfo {author} {\bibfnamefont {L.}~\bibnamefont
  {Senatore}}\ and\ \bibinfo {author} {\bibfnamefont {M.}~\bibnamefont
  {Zaldarriaga}},\ }\href {https://doi.org/10.1007/JHEP12(2010)008} {\bibfield
  {journal} {\bibinfo  {journal} {JHEP}\ }\textbf {\bibinfo {volume} {12}},\
  \bibinfo {pages} {008}},\ \Eprint {https://arxiv.org/abs/0912.2734}
  {arXiv:0912.2734 [hep-th]} \BibitemShut {NoStop}%
\bibitem [{\citenamefont {Marolf}\ and\ \citenamefont
  {Morrison}(2011)}]{Marolf:2010nz}%
  \BibitemOpen
  \bibfield  {author} {\bibinfo {author} {\bibfnamefont {D.}~\bibnamefont
  {Marolf}}\ and\ \bibinfo {author} {\bibfnamefont {I.~A.}\ \bibnamefont
  {Morrison}},\ }\href {https://doi.org/10.1103/PhysRevD.84.044040} {\bibfield
  {journal} {\bibinfo  {journal} {Phys. Rev. D}\ }\textbf {\bibinfo {volume}
  {84}},\ \bibinfo {pages} {044040} (\bibinfo {year} {2011})},\ \Eprint
  {https://arxiv.org/abs/1010.5327} {arXiv:1010.5327 [gr-qc]} \BibitemShut
  {NoStop}%
\bibitem [{\citenamefont {Jatkar}\ \emph {et~al.}(2012)\citenamefont {Jatkar},
  \citenamefont {Leblond},\ and\ \citenamefont {Rajaraman}}]{Jatkar:2011ju}%
  \BibitemOpen
  \bibfield  {author} {\bibinfo {author} {\bibfnamefont {D.~P.}\ \bibnamefont
  {Jatkar}}, \bibinfo {author} {\bibfnamefont {L.}~\bibnamefont {Leblond}},\
  and\ \bibinfo {author} {\bibfnamefont {A.}~\bibnamefont {Rajaraman}},\ }\href
  {https://doi.org/10.1103/PhysRevD.85.024047} {\bibfield  {journal} {\bibinfo
  {journal} {Phys. Rev. D}\ }\textbf {\bibinfo {volume} {85}},\ \bibinfo
  {pages} {024047} (\bibinfo {year} {2012})},\ \Eprint
  {https://arxiv.org/abs/1107.3513} {arXiv:1107.3513 [hep-th]} \BibitemShut
  {NoStop}%
\bibitem [{\citenamefont {Pimentel}\ \emph {et~al.}(2012)\citenamefont
  {Pimentel}, \citenamefont {Senatore},\ and\ \citenamefont
  {Zaldarriaga}}]{Pimentel:2012tw}%
  \BibitemOpen
  \bibfield  {author} {\bibinfo {author} {\bibfnamefont {G.~L.}\ \bibnamefont
  {Pimentel}}, \bibinfo {author} {\bibfnamefont {L.}~\bibnamefont {Senatore}},\
  and\ \bibinfo {author} {\bibfnamefont {M.}~\bibnamefont {Zaldarriaga}},\
  }\href {https://doi.org/10.1007/JHEP07(2012)166} {\bibfield  {journal}
  {\bibinfo  {journal} {JHEP}\ }\textbf {\bibinfo {volume} {07}},\ \bibinfo
  {pages} {166}},\ \Eprint {https://arxiv.org/abs/1203.6651} {arXiv:1203.6651
  [hep-th]} \BibitemShut {NoStop}%
\bibitem [{\citenamefont {Green}\ and\ \citenamefont
  {Premkumar}(2020)}]{Green:2020txs}%
  \BibitemOpen
  \bibfield  {author} {\bibinfo {author} {\bibfnamefont {D.}~\bibnamefont
  {Green}}\ and\ \bibinfo {author} {\bibfnamefont {A.}~\bibnamefont
  {Premkumar}},\ }\href {https://doi.org/10.1007/JHEP04(2020)064} {\bibfield
  {journal} {\bibinfo  {journal} {JHEP}\ }\textbf {\bibinfo {volume} {04}},\
  \bibinfo {pages} {064}},\ \Eprint {https://arxiv.org/abs/2001.05974}
  {arXiv:2001.05974 [hep-th]} \BibitemShut {NoStop}%
\bibitem [{\citenamefont {Cohen}\ and\ \citenamefont
  {Green}(2020)}]{Cohen:2020php}%
  \BibitemOpen
  \bibfield  {author} {\bibinfo {author} {\bibfnamefont {T.}~\bibnamefont
  {Cohen}}\ and\ \bibinfo {author} {\bibfnamefont {D.}~\bibnamefont {Green}},\
  }\href {https://doi.org/10.1007/JHEP12(2020)041} {\bibfield  {journal}
  {\bibinfo  {journal} {JHEP}\ }\textbf {\bibinfo {volume} {12}},\ \bibinfo
  {pages} {041}},\ \Eprint {https://arxiv.org/abs/2007.03693} {arXiv:2007.03693
  [hep-th]} \BibitemShut {NoStop}%
\bibitem [{\citenamefont {Di~Pietro}\ \emph {et~al.}(2023)\citenamefont
  {Di~Pietro}, \citenamefont {Gorbenko},\ and\ \citenamefont
  {Komatsu}}]{DiPietro:2023inn}%
  \BibitemOpen
  \bibfield  {author} {\bibinfo {author} {\bibfnamefont {L.}~\bibnamefont
  {Di~Pietro}}, \bibinfo {author} {\bibfnamefont {V.}~\bibnamefont
  {Gorbenko}},\ and\ \bibinfo {author} {\bibfnamefont {S.}~\bibnamefont
  {Komatsu}},\ }\href@noop {} {\  (\bibinfo {year} {2023})},\ \Eprint
  {https://arxiv.org/abs/2312.17195} {arXiv:2312.17195 [hep-th]} \BibitemShut
  {NoStop}%
\bibitem [{\citenamefont {Heckelbacher}\ \emph {et~al.}(2022)\citenamefont
  {Heckelbacher}, \citenamefont {Sachs}, \citenamefont {Skvortsov},\ and\
  \citenamefont {Vanhove}}]{Heckelbacher:2022hbq}%
  \BibitemOpen
  \bibfield  {author} {\bibinfo {author} {\bibfnamefont {T.}~\bibnamefont
  {Heckelbacher}}, \bibinfo {author} {\bibfnamefont {I.}~\bibnamefont {Sachs}},
  \bibinfo {author} {\bibfnamefont {E.}~\bibnamefont {Skvortsov}},\ and\
  \bibinfo {author} {\bibfnamefont {P.}~\bibnamefont {Vanhove}},\ }\href
  {https://doi.org/10.1007/JHEP08(2022)139} {\bibfield  {journal} {\bibinfo
  {journal} {JHEP}\ }\textbf {\bibinfo {volume} {08}},\ \bibinfo {pages}
  {139}},\ \Eprint {https://arxiv.org/abs/2204.07217} {arXiv:2204.07217
  [hep-th]} \BibitemShut {NoStop}%
\bibitem [{\citenamefont {Chowdhury}\ and\ \citenamefont
  {Singh}(2023)}]{Chowdhury:2023khl}%
  \BibitemOpen
  \bibfield  {author} {\bibinfo {author} {\bibfnamefont {C.}~\bibnamefont
  {Chowdhury}}\ and\ \bibinfo {author} {\bibfnamefont {K.}~\bibnamefont
  {Singh}},\ }\href {https://doi.org/10.1007/JHEP12(2023)109} {\bibfield
  {journal} {\bibinfo  {journal} {JHEP}\ }\textbf {\bibinfo {volume} {12}},\
  \bibinfo {pages} {109}},\ \Eprint {https://arxiv.org/abs/2305.18529}
  {arXiv:2305.18529 [hep-th]} \BibitemShut {NoStop}%
\bibitem [{\citenamefont {Green}\ and\ \citenamefont
  {Gupta}(2025)}]{Green:2024fsz}%
  \BibitemOpen
  \bibfield  {author} {\bibinfo {author} {\bibfnamefont {D.}~\bibnamefont
  {Green}}\ and\ \bibinfo {author} {\bibfnamefont {K.}~\bibnamefont {Gupta}},\
  }\href {https://doi.org/10.1103/9ywk-xccj} {\bibfield  {journal} {\bibinfo
  {journal} {Phys. Rev. D}\ }\textbf {\bibinfo {volume} {112}},\ \bibinfo
  {pages} {043526} (\bibinfo {year} {2025})},\ \Eprint
  {https://arxiv.org/abs/2410.11973} {arXiv:2410.11973 [hep-th]} \BibitemShut
  {NoStop}%
\bibitem [{\citenamefont {Huenupi}\ \emph
  {et~al.}(2024{\natexlab{a}})\citenamefont {Huenupi}, \citenamefont {Hughes},
  \citenamefont {Palma},\ and\ \citenamefont {Sypsas}}]{Huenupi:2024ztu}%
  \BibitemOpen
  \bibfield  {author} {\bibinfo {author} {\bibfnamefont {J.}~\bibnamefont
  {Huenupi}}, \bibinfo {author} {\bibfnamefont {E.}~\bibnamefont {Hughes}},
  \bibinfo {author} {\bibfnamefont {G.~A.}\ \bibnamefont {Palma}},\ and\
  \bibinfo {author} {\bibfnamefont {S.}~\bibnamefont {Sypsas}},\ }\href@noop {}
  {\  (\bibinfo {year} {2024}{\natexlab{a}})},\ \Eprint
  {https://arxiv.org/abs/2412.01891} {arXiv:2412.01891 [hep-th]} \BibitemShut
  {NoStop}%
\bibitem [{\citenamefont {Huenupi}\ \emph
  {et~al.}(2024{\natexlab{b}})\citenamefont {Huenupi}, \citenamefont {Hughes},
  \citenamefont {Palma},\ and\ \citenamefont {Sypsas}}]{Huenupi:2024ksc}%
  \BibitemOpen
  \bibfield  {author} {\bibinfo {author} {\bibfnamefont {J.}~\bibnamefont
  {Huenupi}}, \bibinfo {author} {\bibfnamefont {E.}~\bibnamefont {Hughes}},
  \bibinfo {author} {\bibfnamefont {G.~A.}\ \bibnamefont {Palma}},\ and\
  \bibinfo {author} {\bibfnamefont {S.}~\bibnamefont {Sypsas}},\ }\href
  {https://doi.org/10.1103/PhysRevD.110.123536} {\bibfield  {journal} {\bibinfo
   {journal} {Phys. Rev. D}\ }\textbf {\bibinfo {volume} {110}},\ \bibinfo
  {pages} {123536} (\bibinfo {year} {2024}{\natexlab{b}})},\ \Eprint
  {https://arxiv.org/abs/2406.07610} {arXiv:2406.07610 [hep-th]} \BibitemShut
  {NoStop}%
\bibitem [{\citenamefont {Palma}\ \emph {et~al.}(2025)\citenamefont {Palma},
  \citenamefont {Sypsas},\ and\ \citenamefont {Tapia}}]{Palma:2025oux}%
  \BibitemOpen
  \bibfield  {author} {\bibinfo {author} {\bibfnamefont {G.~A.}\ \bibnamefont
  {Palma}}, \bibinfo {author} {\bibfnamefont {S.}~\bibnamefont {Sypsas}},\ and\
  \bibinfo {author} {\bibfnamefont {D.}~\bibnamefont {Tapia}},\ }\href@noop {}
  {\  (\bibinfo {year} {2025})},\ \Eprint {https://arxiv.org/abs/2507.21310}
  {arXiv:2507.21310 [hep-th]} \BibitemShut {NoStop}%
\bibitem [{\citenamefont {Benincasa}\ \emph {et~al.}(2025)\citenamefont
  {Benincasa}, \citenamefont {Brunello}, \citenamefont {Mandal}, \citenamefont
  {Mastrolia},\ and\ \citenamefont {Vaz{\~a}o}}]{Benincasa:2024ptf}%
  \BibitemOpen
  \bibfield  {author} {\bibinfo {author} {\bibfnamefont {P.}~\bibnamefont
  {Benincasa}}, \bibinfo {author} {\bibfnamefont {G.}~\bibnamefont {Brunello}},
  \bibinfo {author} {\bibfnamefont {M.~K.}\ \bibnamefont {Mandal}}, \bibinfo
  {author} {\bibfnamefont {P.}~\bibnamefont {Mastrolia}},\ and\ \bibinfo
  {author} {\bibfnamefont {F.}~\bibnamefont {Vaz{\~a}o}},\ }\href
  {https://doi.org/10.1103/PhysRevD.111.085016} {\bibfield  {journal} {\bibinfo
   {journal} {Phys. Rev. D}\ }\textbf {\bibinfo {volume} {111}},\ \bibinfo
  {pages} {085016} (\bibinfo {year} {2025})},\ \Eprint
  {https://arxiv.org/abs/2408.16386} {arXiv:2408.16386 [hep-th]} \BibitemShut
  {NoStop}%
\bibitem [{\citenamefont {Braglia}\ and\ \citenamefont
  {Pinol}(2025{\natexlab{a}})}]{Braglia:2025cee}%
  \BibitemOpen
  \bibfield  {author} {\bibinfo {author} {\bibfnamefont {M.}~\bibnamefont
  {Braglia}}\ and\ \bibinfo {author} {\bibfnamefont {L.}~\bibnamefont
  {Pinol}},\ }\href@noop {} {\  (\bibinfo {year} {2025}{\natexlab{a}})},\
  \Eprint {https://arxiv.org/abs/2504.07926} {arXiv:2504.07926 [astro-ph.CO]}
  \BibitemShut {NoStop}%
\bibitem [{\citenamefont {Nowinski}\ and\ \citenamefont
  {Sachs}(2025)}]{Nowinski:2025cvw}%
  \BibitemOpen
  \bibfield  {author} {\bibinfo {author} {\bibfnamefont {M.}~\bibnamefont
  {Nowinski}}\ and\ \bibinfo {author} {\bibfnamefont {I.}~\bibnamefont
  {Sachs}},\ }\href@noop {} {\  (\bibinfo {year} {2025})},\ \Eprint
  {https://arxiv.org/abs/2507.21224} {arXiv:2507.21224 [hep-th]} \BibitemShut
  {NoStop}%
\bibitem [{\citenamefont {Bhowmick}\ \emph {et~al.}(2024)\citenamefont
  {Bhowmick}, \citenamefont {Ghosh},\ and\ \citenamefont
  {Ullah}}]{Bhowmick:2024kld}%
  \BibitemOpen
  \bibfield  {author} {\bibinfo {author} {\bibfnamefont {S.}~\bibnamefont
  {Bhowmick}}, \bibinfo {author} {\bibfnamefont {D.}~\bibnamefont {Ghosh}},\
  and\ \bibinfo {author} {\bibfnamefont {F.}~\bibnamefont {Ullah}},\ }\href
  {https://doi.org/10.1007/JHEP10(2024)057} {\bibfield  {journal} {\bibinfo
  {journal} {JHEP}\ }\textbf {\bibinfo {volume} {10}},\ \bibinfo {pages}
  {057}},\ \Eprint {https://arxiv.org/abs/2405.10374} {arXiv:2405.10374
  [hep-th]} \BibitemShut {NoStop}%
\bibitem [{\citenamefont {Bhowmick}\ \emph {et~al.}(2026)\citenamefont
  {Bhowmick}, \citenamefont {Lee}, \citenamefont {Ghosh},\ and\ \citenamefont
  {Ullah}}]{Bhowmick:2025mxh}%
  \BibitemOpen
  \bibfield  {author} {\bibinfo {author} {\bibfnamefont {S.}~\bibnamefont
  {Bhowmick}}, \bibinfo {author} {\bibfnamefont {M.~H.~G.}\ \bibnamefont
  {Lee}}, \bibinfo {author} {\bibfnamefont {D.}~\bibnamefont {Ghosh}},\ and\
  \bibinfo {author} {\bibfnamefont {F.}~\bibnamefont {Ullah}},\ }\href
  {https://doi.org/10.1007/JHEP02(2026)116} {\bibfield  {journal} {\bibinfo
  {journal} {JHEP}\ }\textbf {\bibinfo {volume} {02}},\ \bibinfo {pages}
  {116}},\ \Eprint {https://arxiv.org/abs/2503.21880} {arXiv:2503.21880
  [hep-th]} \BibitemShut {NoStop}%
\bibitem [{\citenamefont {Chowdhury}\ \emph {et~al.}(2025)\citenamefont
  {Chowdhury}, \citenamefont {Lipstein}, \citenamefont {Mei}, \citenamefont
  {Sachs},\ and\ \citenamefont {Vanhove}}]{Chowdhury:2023arc}%
  \BibitemOpen
  \bibfield  {author} {\bibinfo {author} {\bibfnamefont {C.}~\bibnamefont
  {Chowdhury}}, \bibinfo {author} {\bibfnamefont {A.}~\bibnamefont {Lipstein}},
  \bibinfo {author} {\bibfnamefont {J.}~\bibnamefont {Mei}}, \bibinfo {author}
  {\bibfnamefont {I.}~\bibnamefont {Sachs}},\ and\ \bibinfo {author}
  {\bibfnamefont {P.}~\bibnamefont {Vanhove}},\ }\href
  {https://doi.org/10.1007/JHEP03(2025)007} {\bibfield  {journal} {\bibinfo
  {journal} {JHEP}\ }\textbf {\bibinfo {volume} {03}},\ \bibinfo {pages}
  {007}},\ \Eprint {https://arxiv.org/abs/2312.13803} {arXiv:2312.13803
  [hep-th]} \BibitemShut {NoStop}%
\bibitem [{\citenamefont {Chowdhury}\ \emph {et~al.}(2024)\citenamefont
  {Chowdhury}, \citenamefont {Chowdhury}, \citenamefont {Moga},\ and\
  \citenamefont {Singh}}]{Chowdhury:2024snc}%
  \BibitemOpen
  \bibfield  {author} {\bibinfo {author} {\bibfnamefont {C.}~\bibnamefont
  {Chowdhury}}, \bibinfo {author} {\bibfnamefont {P.}~\bibnamefont
  {Chowdhury}}, \bibinfo {author} {\bibfnamefont {R.~N.}\ \bibnamefont
  {Moga}},\ and\ \bibinfo {author} {\bibfnamefont {K.}~\bibnamefont {Singh}},\
  }\href {https://doi.org/10.1007/JHEP10(2024)202} {\bibfield  {journal}
  {\bibinfo  {journal} {JHEP}\ }\textbf {\bibinfo {volume} {10}},\ \bibinfo
  {pages} {202}},\ \Eprint {https://arxiv.org/abs/2408.00074} {arXiv:2408.00074
  [hep-th]} \BibitemShut {NoStop}%
\bibitem [{\citenamefont {Premkumar}(2024)}]{Premkumar:2021mlz}%
  \BibitemOpen
  \bibfield  {author} {\bibinfo {author} {\bibfnamefont {A.}~\bibnamefont
  {Premkumar}},\ }\href {https://doi.org/10.1103/PhysRevD.109.045003}
  {\bibfield  {journal} {\bibinfo  {journal} {Phys. Rev. D}\ }\textbf {\bibinfo
  {volume} {109}},\ \bibinfo {pages} {045003} (\bibinfo {year} {2024})},\
  \Eprint {https://arxiv.org/abs/2110.12504} {arXiv:2110.12504 [hep-th]}
  \BibitemShut {NoStop}%
\bibitem [{\citenamefont {Melville}\ and\ \citenamefont
  {Pajer}(2021)}]{Melville:2021lst}%
  \BibitemOpen
  \bibfield  {author} {\bibinfo {author} {\bibfnamefont {S.}~\bibnamefont
  {Melville}}\ and\ \bibinfo {author} {\bibfnamefont {E.}~\bibnamefont
  {Pajer}},\ }\href {https://doi.org/10.1007/JHEP05(2021)249} {\bibfield
  {journal} {\bibinfo  {journal} {JHEP}\ }\textbf {\bibinfo {volume} {05}},\
  \bibinfo {pages} {249}},\ \Eprint {https://arxiv.org/abs/2103.09832}
  {arXiv:2103.09832 [hep-th]} \BibitemShut {NoStop}%
\bibitem [{\citenamefont {C\'espedes}\ \emph {et~al.}(2024)\citenamefont
  {C\'espedes}, \citenamefont {Davis},\ and\ \citenamefont
  {Wang}}]{Cespedes:2023aal}%
  \BibitemOpen
  \bibfield  {author} {\bibinfo {author} {\bibfnamefont {S.}~\bibnamefont
  {C\'espedes}}, \bibinfo {author} {\bibfnamefont {A.-C.}\ \bibnamefont
  {Davis}},\ and\ \bibinfo {author} {\bibfnamefont {D.-G.}\ \bibnamefont
  {Wang}},\ }\href {https://doi.org/10.1007/JHEP04(2024)004} {\bibfield
  {journal} {\bibinfo  {journal} {JHEP}\ }\textbf {\bibinfo {volume} {04}},\
  \bibinfo {pages} {004}},\ \Eprint {https://arxiv.org/abs/2311.17990}
  {arXiv:2311.17990 [hep-th]} \BibitemShut {NoStop}%
\bibitem [{\citenamefont {Qin}\ and\ \citenamefont
  {Xianyu}(2023)}]{Qin:2023bjk}%
  \BibitemOpen
  \bibfield  {author} {\bibinfo {author} {\bibfnamefont {Z.}~\bibnamefont
  {Qin}}\ and\ \bibinfo {author} {\bibfnamefont {Z.-Z.}\ \bibnamefont
  {Xianyu}},\ }\href {https://doi.org/10.1007/JHEP09(2023)116} {\bibfield
  {journal} {\bibinfo  {journal} {JHEP}\ }\textbf {\bibinfo {volume} {09}},\
  \bibinfo {pages} {116}},\ \Eprint {https://arxiv.org/abs/2304.13295}
  {arXiv:2304.13295 [hep-th]} \BibitemShut {NoStop}%
\bibitem [{\citenamefont {Qin}\ and\ \citenamefont
  {Xianyu}(2024)}]{Qin:2023nhv}%
  \BibitemOpen
  \bibfield  {author} {\bibinfo {author} {\bibfnamefont {Z.}~\bibnamefont
  {Qin}}\ and\ \bibinfo {author} {\bibfnamefont {Z.-Z.}\ \bibnamefont
  {Xianyu}},\ }\href {https://doi.org/10.1007/JHEP01(2024)168} {\bibfield
  {journal} {\bibinfo  {journal} {JHEP}\ }\textbf {\bibinfo {volume} {01}},\
  \bibinfo {pages} {168}},\ \Eprint {https://arxiv.org/abs/2308.14802}
  {arXiv:2308.14802 [hep-th]} \BibitemShut {NoStop}%
\bibitem [{\citenamefont {Qin}(2025)}]{Qin:2024gtr}%
  \BibitemOpen
  \bibfield  {author} {\bibinfo {author} {\bibfnamefont {Z.}~\bibnamefont
  {Qin}},\ }\href {https://doi.org/10.1007/JHEP03(2025)051} {\bibfield
  {journal} {\bibinfo  {journal} {JHEP}\ }\textbf {\bibinfo {volume} {03}},\
  \bibinfo {pages} {051}},\ \Eprint {https://arxiv.org/abs/2411.13636}
  {arXiv:2411.13636 [hep-th]} \BibitemShut {NoStop}%
\bibitem [{\citenamefont {Liu}\ \emph {et~al.}(2025)\citenamefont {Liu},
  \citenamefont {Qin},\ and\ \citenamefont {Xianyu}}]{Liu:2024xyi}%
  \BibitemOpen
  \bibfield  {author} {\bibinfo {author} {\bibfnamefont {H.}~\bibnamefont
  {Liu}}, \bibinfo {author} {\bibfnamefont {Z.}~\bibnamefont {Qin}},\ and\
  \bibinfo {author} {\bibfnamefont {Z.-Z.}\ \bibnamefont {Xianyu}},\ }\href
  {https://doi.org/10.1007/JHEP02(2025)101} {\bibfield  {journal} {\bibinfo
  {journal} {JHEP}\ }\textbf {\bibinfo {volume} {02}},\ \bibinfo {pages}
  {101}},\ \Eprint {https://arxiv.org/abs/2407.12299} {arXiv:2407.12299
  [hep-th]} \BibitemShut {NoStop}%
\bibitem [{\citenamefont {Baumann}\ \emph {et~al.}(2025)\citenamefont
  {Baumann}, \citenamefont {Goodhew},\ and\ \citenamefont
  {Lee}}]{Baumann:2024mvm}%
  \BibitemOpen
  \bibfield  {author} {\bibinfo {author} {\bibfnamefont {D.}~\bibnamefont
  {Baumann}}, \bibinfo {author} {\bibfnamefont {H.}~\bibnamefont {Goodhew}},\
  and\ \bibinfo {author} {\bibfnamefont {H.}~\bibnamefont {Lee}},\ }\href
  {https://doi.org/10.1007/JHEP07(2025)131} {\bibfield  {journal} {\bibinfo
  {journal} {JHEP}\ }\textbf {\bibinfo {volume} {07}},\ \bibinfo {pages}
  {131}},\ \Eprint {https://arxiv.org/abs/2410.17994} {arXiv:2410.17994
  [hep-th]} \BibitemShut {NoStop}%
\bibitem [{\citenamefont {Ansari}\ \emph {et~al.}(2025)\citenamefont {Ansari},
  \citenamefont {Bhowmick},\ and\ \citenamefont {Ghosh}}]{Ansari:2025nng}%
  \BibitemOpen
  \bibfield  {author} {\bibinfo {author} {\bibfnamefont {E.}~\bibnamefont
  {Ansari}}, \bibinfo {author} {\bibfnamefont {S.}~\bibnamefont {Bhowmick}},\
  and\ \bibinfo {author} {\bibfnamefont {D.}~\bibnamefont {Ghosh}},\
  }\href@noop {} {\  (\bibinfo {year} {2025})},\ \Eprint
  {https://arxiv.org/abs/2512.11040} {arXiv:2512.11040 [hep-th]} \BibitemShut
  {NoStop}%
\bibitem [{\citenamefont {Pimentel}\ and\ \citenamefont
  {Westerdijk}(2026)}]{Pimentel:2026kqc}%
  \BibitemOpen
  \bibfield  {author} {\bibinfo {author} {\bibfnamefont {G.~L.}\ \bibnamefont
  {Pimentel}}\ and\ \bibinfo {author} {\bibfnamefont {T.}~\bibnamefont
  {Westerdijk}},\ }\href@noop {} {\  (\bibinfo {year} {2026})},\ \Eprint
  {https://arxiv.org/abs/2601.00952} {arXiv:2601.00952 [hep-th]} \BibitemShut
  {NoStop}%
\bibitem [{\citenamefont {Gr{\"a}fe}\ and\ \citenamefont
  {Sachs}(2026)}]{Grafe:2026qsm}%
  \BibitemOpen
  \bibfield  {author} {\bibinfo {author} {\bibfnamefont {J.}~\bibnamefont
  {Gr{\"a}fe}}\ and\ \bibinfo {author} {\bibfnamefont {I.}~\bibnamefont
  {Sachs}},\ }\href@noop {} {\  (\bibinfo {year} {2026})},\ \Eprint
  {https://arxiv.org/abs/2602.09977} {arXiv:2602.09977 [hep-th]} \BibitemShut
  {NoStop}%
\bibitem [{\citenamefont {Chowdhury}\ \emph {et~al.}(2026)\citenamefont
  {Chowdhury}, \citenamefont {Jazayeri}, \citenamefont {Lipstein},
  \citenamefont {Marshall}, \citenamefont {Mei},\ and\ \citenamefont
  {Sachs}}]{Chowdhury:2026upp}%
  \BibitemOpen
  \bibfield  {author} {\bibinfo {author} {\bibfnamefont {C.}~\bibnamefont
  {Chowdhury}}, \bibinfo {author} {\bibfnamefont {S.}~\bibnamefont {Jazayeri}},
  \bibinfo {author} {\bibfnamefont {A.}~\bibnamefont {Lipstein}}, \bibinfo
  {author} {\bibfnamefont {J.}~\bibnamefont {Marshall}}, \bibinfo {author}
  {\bibfnamefont {J.}~\bibnamefont {Mei}},\ and\ \bibinfo {author}
  {\bibfnamefont {I.}~\bibnamefont {Sachs}},\ }\href@noop {} {\  (\bibinfo
  {year} {2026})},\ \Eprint {https://arxiv.org/abs/2602.03841}
  {arXiv:2602.03841 [hep-th]} \BibitemShut {NoStop}%
\bibitem [{\citenamefont {Chen}\ \emph {et~al.}(2016)\citenamefont {Chen},
  \citenamefont {Wang},\ and\ \citenamefont {Xianyu}}]{Chen:2016nrs}%
  \BibitemOpen
  \bibfield  {author} {\bibinfo {author} {\bibfnamefont {X.}~\bibnamefont
  {Chen}}, \bibinfo {author} {\bibfnamefont {Y.}~\bibnamefont {Wang}},\ and\
  \bibinfo {author} {\bibfnamefont {Z.-Z.}\ \bibnamefont {Xianyu}},\ }\href
  {https://doi.org/10.1007/JHEP08(2016)051} {\bibfield  {journal} {\bibinfo
  {journal} {JHEP}\ }\textbf {\bibinfo {volume} {08}},\ \bibinfo {pages}
  {051}},\ \Eprint {https://arxiv.org/abs/1604.07841} {arXiv:1604.07841
  [hep-th]} \BibitemShut {NoStop}%
\bibitem [{\citenamefont {Wang}\ \emph {et~al.}(2022)\citenamefont {Wang},
  \citenamefont {Xianyu},\ and\ \citenamefont {Zhong}}]{Wang:2021qez}%
  \BibitemOpen
  \bibfield  {author} {\bibinfo {author} {\bibfnamefont {L.-T.}\ \bibnamefont
  {Wang}}, \bibinfo {author} {\bibfnamefont {Z.-Z.}\ \bibnamefont {Xianyu}},\
  and\ \bibinfo {author} {\bibfnamefont {Y.-M.}\ \bibnamefont {Zhong}},\ }\href
  {https://doi.org/10.1007/JHEP02(2022)085} {\bibfield  {journal} {\bibinfo
  {journal} {JHEP}\ }\textbf {\bibinfo {volume} {02}},\ \bibinfo {pages}
  {085}},\ \Eprint {https://arxiv.org/abs/2109.14635} {arXiv:2109.14635
  [hep-ph]} \BibitemShut {NoStop}%
\bibitem [{\citenamefont {Xianyu}\ and\ \citenamefont
  {Zhang}(2023)}]{Xianyu:2022jwk}%
  \BibitemOpen
  \bibfield  {author} {\bibinfo {author} {\bibfnamefont {Z.-Z.}\ \bibnamefont
  {Xianyu}}\ and\ \bibinfo {author} {\bibfnamefont {H.}~\bibnamefont {Zhang}},\
  }\href {https://doi.org/10.1007/JHEP04(2023)103} {\bibfield  {journal}
  {\bibinfo  {journal} {JHEP}\ }\textbf {\bibinfo {volume} {04}},\ \bibinfo
  {pages} {103}},\ \Eprint {https://arxiv.org/abs/2211.03810} {arXiv:2211.03810
  [hep-th]} \BibitemShut {NoStop}%
\bibitem [{\citenamefont {Bodas}\ \emph {et~al.}(2025)\citenamefont {Bodas},
  \citenamefont {Broadberry}, \citenamefont {Sundrum},\ and\ \citenamefont
  {Xu}}]{Bodas:2025wuk}%
  \BibitemOpen
  \bibfield  {author} {\bibinfo {author} {\bibfnamefont {A.}~\bibnamefont
  {Bodas}}, \bibinfo {author} {\bibfnamefont {E.}~\bibnamefont {Broadberry}},
  \bibinfo {author} {\bibfnamefont {R.}~\bibnamefont {Sundrum}},\ and\ \bibinfo
  {author} {\bibfnamefont {Z.}~\bibnamefont {Xu}},\ }\href@noop {} {\
  (\bibinfo {year} {2025})},\ \Eprint {https://arxiv.org/abs/2507.22978}
  {arXiv:2507.22978 [hep-ph]} \BibitemShut {NoStop}%
\bibitem [{\citenamefont {Kristiano}\ and\ \citenamefont
  {Yokoyama}(2024)}]{Kristiano:2023scm}%
  \BibitemOpen
  \bibfield  {author} {\bibinfo {author} {\bibfnamefont {J.}~\bibnamefont
  {Kristiano}}\ and\ \bibinfo {author} {\bibfnamefont {J.}~\bibnamefont
  {Yokoyama}},\ }\href {https://doi.org/10.1103/PhysRevD.109.103541} {\bibfield
   {journal} {\bibinfo  {journal} {Phys. Rev. D}\ }\textbf {\bibinfo {volume}
  {109}},\ \bibinfo {pages} {103541} (\bibinfo {year} {2024})},\ \Eprint
  {https://arxiv.org/abs/2303.00341} {arXiv:2303.00341 [hep-th]} \BibitemShut
  {NoStop}%
\bibitem [{\citenamefont {Firouzjahi}\ and\ \citenamefont
  {Riotto}(2024)}]{Firouzjahi:2023ahg}%
  \BibitemOpen
  \bibfield  {author} {\bibinfo {author} {\bibfnamefont {H.}~\bibnamefont
  {Firouzjahi}}\ and\ \bibinfo {author} {\bibfnamefont {A.}~\bibnamefont
  {Riotto}},\ }\href {https://doi.org/10.1088/1475-7516/2024/02/021} {\bibfield
   {journal} {\bibinfo  {journal} {JCAP}\ }\textbf {\bibinfo {volume} {02}},\
  \bibinfo {pages} {021}},\ \Eprint {https://arxiv.org/abs/2304.07801}
  {arXiv:2304.07801 [astro-ph.CO]} \BibitemShut {NoStop}%
\bibitem [{\citenamefont {Beneke}\ \emph {et~al.}(2024)\citenamefont {Beneke},
  \citenamefont {Hager},\ and\ \citenamefont {Sanfilippo}}]{Beneke:2023wmt}%
  \BibitemOpen
  \bibfield  {author} {\bibinfo {author} {\bibfnamefont {M.}~\bibnamefont
  {Beneke}}, \bibinfo {author} {\bibfnamefont {P.}~\bibnamefont {Hager}},\ and\
  \bibinfo {author} {\bibfnamefont {A.~F.}\ \bibnamefont {Sanfilippo}},\ }\href
  {https://doi.org/10.1007/JHEP04(2024)006} {\bibfield  {journal} {\bibinfo
  {journal} {JHEP}\ }\textbf {\bibinfo {volume} {04}},\ \bibinfo {pages}
  {006}},\ \Eprint {https://arxiv.org/abs/2312.06766} {arXiv:2312.06766
  [hep-th]} \BibitemShut {NoStop}%
\bibitem [{\citenamefont {Fumagalli}(2025)}]{Fumagalli:2023zzl}%
  \BibitemOpen
  \bibfield  {author} {\bibinfo {author} {\bibfnamefont {J.}~\bibnamefont
  {Fumagalli}},\ }\href {https://doi.org/10.1007/JHEP05(2025)162} {\bibfield
  {journal} {\bibinfo  {journal} {JHEP}\ }\textbf {\bibinfo {volume} {05}},\
  \bibinfo {pages} {162}},\ \Eprint {https://arxiv.org/abs/2305.19263}
  {arXiv:2305.19263 [astro-ph.CO]} \BibitemShut {NoStop}%
\bibitem [{\citenamefont {Choudhury}\ \emph {et~al.}(2023)\citenamefont
  {Choudhury}, \citenamefont {Panda},\ and\ \citenamefont
  {Sami}}]{Choudhury:2023rks}%
  \BibitemOpen
  \bibfield  {author} {\bibinfo {author} {\bibfnamefont {S.}~\bibnamefont
  {Choudhury}}, \bibinfo {author} {\bibfnamefont {S.}~\bibnamefont {Panda}},\
  and\ \bibinfo {author} {\bibfnamefont {M.}~\bibnamefont {Sami}},\ }\href
  {https://doi.org/10.1088/1475-7516/2023/11/066} {\bibfield  {journal}
  {\bibinfo  {journal} {JCAP}\ }\textbf {\bibinfo {volume} {11}},\ \bibinfo
  {pages} {066}},\ \Eprint {https://arxiv.org/abs/2303.06066} {arXiv:2303.06066
  [astro-ph.CO]} \BibitemShut {NoStop}%
\bibitem [{\citenamefont {Firouzjahi}(2023)}]{Firouzjahi:2023aum}%
  \BibitemOpen
  \bibfield  {author} {\bibinfo {author} {\bibfnamefont {H.}~\bibnamefont
  {Firouzjahi}},\ }\href {https://doi.org/10.1088/1475-7516/2023/10/006}
  {\bibfield  {journal} {\bibinfo  {journal} {JCAP}\ }\textbf {\bibinfo
  {volume} {10}},\ \bibinfo {pages} {006}},\ \Eprint
  {https://arxiv.org/abs/2303.12025} {arXiv:2303.12025 [astro-ph.CO]}
  \BibitemShut {NoStop}%
\bibitem [{\citenamefont {Frolovsky}\ and\ \citenamefont
  {Ketov}(2025)}]{Frolovsky:2025qre}%
  \BibitemOpen
  \bibfield  {author} {\bibinfo {author} {\bibfnamefont {D.}~\bibnamefont
  {Frolovsky}}\ and\ \bibinfo {author} {\bibfnamefont {S.~V.}\ \bibnamefont
  {Ketov}},\ }\href {https://doi.org/10.1103/PhysRevD.111.083533} {\bibfield
  {journal} {\bibinfo  {journal} {Phys. Rev. D}\ }\textbf {\bibinfo {volume}
  {111}},\ \bibinfo {pages} {083533} (\bibinfo {year} {2025})},\ \Eprint
  {https://arxiv.org/abs/2502.00628} {arXiv:2502.00628 [gr-qc]} \BibitemShut
  {NoStop}%
\bibitem [{\citenamefont {Inomata}(2025{\natexlab{a}})}]{Inomata:2025bqw}%
  \BibitemOpen
  \bibfield  {author} {\bibinfo {author} {\bibfnamefont {K.}~\bibnamefont
  {Inomata}},\ }\href {https://doi.org/10.1103/PhysRevD.111.103504} {\bibfield
  {journal} {\bibinfo  {journal} {Phys. Rev. D}\ }\textbf {\bibinfo {volume}
  {111}},\ \bibinfo {pages} {103504} (\bibinfo {year} {2025}{\natexlab{a}})},\
  \Eprint {https://arxiv.org/abs/2502.08707} {arXiv:2502.08707 [astro-ph.CO]}
  \BibitemShut {NoStop}%
\bibitem [{\citenamefont {Fang}\ \emph {et~al.}(2025)\citenamefont {Fang},
  \citenamefont {Lyu}, \citenamefont {Chen},\ and\ \citenamefont
  {Guo}}]{Fang:2025vhi}%
  \BibitemOpen
  \bibfield  {author} {\bibinfo {author} {\bibfnamefont {C.-J.}\ \bibnamefont
  {Fang}}, \bibinfo {author} {\bibfnamefont {Z.-H.}\ \bibnamefont {Lyu}},
  \bibinfo {author} {\bibfnamefont {C.}~\bibnamefont {Chen}},\ and\ \bibinfo
  {author} {\bibfnamefont {Z.-K.}\ \bibnamefont {Guo}},\ }\href
  {https://doi.org/10.1103/nkrq-3d39} {\bibfield  {journal} {\bibinfo
  {journal} {Phys. Rev. D}\ }\textbf {\bibinfo {volume} {112}},\ \bibinfo
  {pages} {023547} (\bibinfo {year} {2025})},\ \Eprint
  {https://arxiv.org/abs/2502.09555} {arXiv:2502.09555 [gr-qc]} \BibitemShut
  {NoStop}%
\bibitem [{\citenamefont {Inomata}(2025{\natexlab{b}})}]{Inomata:2025pqa}%
  \BibitemOpen
  \bibfield  {author} {\bibinfo {author} {\bibfnamefont {K.}~\bibnamefont
  {Inomata}},\ }\href {https://doi.org/10.1103/r8bh-s48f} {\bibfield  {journal}
  {\bibinfo  {journal} {Phys. Rev. D}\ }\textbf {\bibinfo {volume} {111}},\
  \bibinfo {pages} {123517} (\bibinfo {year} {2025}{\natexlab{b}})},\ \Eprint
  {https://arxiv.org/abs/2502.12112} {arXiv:2502.12112 [astro-ph.CO]}
  \BibitemShut {NoStop}%
\bibitem [{\citenamefont {Braglia}\ and\ \citenamefont
  {Pinol}(2025{\natexlab{b}})}]{Braglia:2025qrb}%
  \BibitemOpen
  \bibfield  {author} {\bibinfo {author} {\bibfnamefont {M.}~\bibnamefont
  {Braglia}}\ and\ \bibinfo {author} {\bibfnamefont {L.}~\bibnamefont
  {Pinol}},\ }\href@noop {} {\  (\bibinfo {year} {2025}{\natexlab{b}})},\
  \Eprint {https://arxiv.org/abs/2504.13136} {arXiv:2504.13136 [astro-ph.CO]}
  \BibitemShut {NoStop}%
\bibitem [{\citenamefont {Ema}\ \emph {et~al.}(2025)\citenamefont {Ema},
  \citenamefont {Hong}, \citenamefont {Jinno},\ and\ \citenamefont
  {Mukaida}}]{Ema:2025ftj}%
  \BibitemOpen
  \bibfield  {author} {\bibinfo {author} {\bibfnamefont {Y.}~\bibnamefont
  {Ema}}, \bibinfo {author} {\bibfnamefont {M.}~\bibnamefont {Hong}}, \bibinfo
  {author} {\bibfnamefont {R.}~\bibnamefont {Jinno}},\ and\ \bibinfo {author}
  {\bibfnamefont {K.}~\bibnamefont {Mukaida}},\ }\href@noop {} {\  (\bibinfo
  {year} {2025})},\ \Eprint {https://arxiv.org/abs/2506.15780}
  {arXiv:2506.15780 [astro-ph.CO]} \BibitemShut {NoStop}%
\bibitem [{\citenamefont {Wang}\ \emph {et~al.}(2025)\citenamefont {Wang},
  \citenamefont {Wang}, \citenamefont {Wang},\ and\ \citenamefont
  {Yu}}]{Wang:2025qfh}%
  \BibitemOpen
  \bibfield  {author} {\bibinfo {author} {\bibfnamefont {D.}~\bibnamefont
  {Wang}}, \bibinfo {author} {\bibfnamefont {X.}~\bibnamefont {Wang}}, \bibinfo
  {author} {\bibfnamefont {Y.}~\bibnamefont {Wang}},\ and\ \bibinfo {author}
  {\bibfnamefont {W.}~\bibnamefont {Yu}},\ }\href@noop {} {\  (\bibinfo {year}
  {2025})},\ \Eprint {https://arxiv.org/abs/2508.12856} {arXiv:2508.12856
  [hep-th]} \BibitemShut {NoStop}%
\bibitem [{\citenamefont {Maru}\ and\ \citenamefont
  {Okawa}(2021)}]{Maru:2021ezc}%
  \BibitemOpen
  \bibfield  {author} {\bibinfo {author} {\bibfnamefont {N.}~\bibnamefont
  {Maru}}\ and\ \bibinfo {author} {\bibfnamefont {A.}~\bibnamefont {Okawa}},\
  }\href@noop {} {\  (\bibinfo {year} {2021})},\ \Eprint
  {https://arxiv.org/abs/2101.10634} {arXiv:2101.10634 [hep-ph]} \BibitemShut
  {NoStop}%
\bibitem [{\citenamefont {Niu}\ \emph {et~al.}(2023)\citenamefont {Niu},
  \citenamefont {Rahat}, \citenamefont {Srinivasan},\ and\ \citenamefont
  {Xue}}]{Niu:2022fki}%
  \BibitemOpen
  \bibfield  {author} {\bibinfo {author} {\bibfnamefont {X.}~\bibnamefont
  {Niu}}, \bibinfo {author} {\bibfnamefont {M.~H.}\ \bibnamefont {Rahat}},
  \bibinfo {author} {\bibfnamefont {K.}~\bibnamefont {Srinivasan}},\ and\
  \bibinfo {author} {\bibfnamefont {W.}~\bibnamefont {Xue}},\ }\href
  {https://doi.org/10.1088/1475-7516/2023/05/018} {\bibfield  {journal}
  {\bibinfo  {journal} {JCAP}\ }\textbf {\bibinfo {volume} {05}},\ \bibinfo
  {pages} {018}},\ \Eprint {https://arxiv.org/abs/2211.14324} {arXiv:2211.14324
  [hep-ph]} \BibitemShut {NoStop}%
\bibitem [{\citenamefont {Fujita}\ \emph {et~al.}(2024)\citenamefont {Fujita},
  \citenamefont {Murata}, \citenamefont {Obata},\ and\ \citenamefont
  {Shiraishi}}]{Fujita:2023inz}%
  \BibitemOpen
  \bibfield  {author} {\bibinfo {author} {\bibfnamefont {T.}~\bibnamefont
  {Fujita}}, \bibinfo {author} {\bibfnamefont {T.}~\bibnamefont {Murata}},
  \bibinfo {author} {\bibfnamefont {I.}~\bibnamefont {Obata}},\ and\ \bibinfo
  {author} {\bibfnamefont {M.}~\bibnamefont {Shiraishi}},\ }\href
  {https://doi.org/10.1088/1475-7516/2024/05/127} {\bibfield  {journal}
  {\bibinfo  {journal} {JCAP}\ }\textbf {\bibinfo {volume} {05}},\ \bibinfo
  {pages} {127}},\ \Eprint {https://arxiv.org/abs/2310.03551} {arXiv:2310.03551
  [astro-ph.CO]} \BibitemShut {NoStop}%
\bibitem [{\citenamefont {Reinhard}\ \emph {et~al.}(2024)\citenamefont
  {Reinhard}, \citenamefont {Slepian}, \citenamefont {Hou},\ and\ \citenamefont
  {Greco}}]{Reinhard:2024evr}%
  \BibitemOpen
  \bibfield  {author} {\bibinfo {author} {\bibfnamefont {M.}~\bibnamefont
  {Reinhard}}, \bibinfo {author} {\bibfnamefont {Z.}~\bibnamefont {Slepian}},
  \bibinfo {author} {\bibfnamefont {J.}~\bibnamefont {Hou}},\ and\ \bibinfo
  {author} {\bibfnamefont {A.}~\bibnamefont {Greco}},\ }\href@noop {} {\
  (\bibinfo {year} {2024})},\ \Eprint {https://arxiv.org/abs/2412.16037}
  {arXiv:2412.16037 [astro-ph.CO]} \BibitemShut {NoStop}%
\bibitem [{\citenamefont {Garcia-Saenz}\ \emph {et~al.}(2025)\citenamefont
  {Garcia-Saenz}, \citenamefont {Lu},\ and\ \citenamefont
  {Renaux-Petel}}]{Garcia-Saenz:2025jis}%
  \BibitemOpen
  \bibfield  {author} {\bibinfo {author} {\bibfnamefont {S.}~\bibnamefont
  {Garcia-Saenz}}, \bibinfo {author} {\bibfnamefont {Y.}~\bibnamefont {Lu}},\
  and\ \bibinfo {author} {\bibfnamefont {S.}~\bibnamefont {Renaux-Petel}},\
  }\href@noop {} {\  (\bibinfo {year} {2025})},\ \Eprint
  {https://arxiv.org/abs/2503.18516} {arXiv:2503.18516 [hep-th]} \BibitemShut
  {NoStop}%
\bibitem [{\citenamefont {Caron-Huot}\ and\ \citenamefont
  {Wilhelm}(2016)}]{Caron_Huot_2016}%
  \BibitemOpen
  \bibfield  {author} {\bibinfo {author} {\bibfnamefont {S.}~\bibnamefont
  {Caron-Huot}}\ and\ \bibinfo {author} {\bibfnamefont {M.}~\bibnamefont
  {Wilhelm}},\ }\bibfield  {journal} {\bibinfo  {journal} {Journal of High
  Energy Physics}\ }\textbf {\bibinfo {volume} {2016}},\ \href
  {https://doi.org/10.1007/jhep12(2016)010} {10.1007/jhep12(2016)010} (\bibinfo
  {year} {2016})\BibitemShut {NoStop}%
\bibitem [{\citenamefont {Chavda}\ \emph
  {et~al.}(2025{\natexlab{a}})\citenamefont {Chavda}, \citenamefont
  {McLoughlin}, \citenamefont {Mizera},\ and\ \citenamefont
  {Staunton}}]{Chavda:2025aqm}%
  \BibitemOpen
  \bibfield  {author} {\bibinfo {author} {\bibfnamefont {A.}~\bibnamefont
  {Chavda}}, \bibinfo {author} {\bibfnamefont {D.}~\bibnamefont {McLoughlin}},
  \bibinfo {author} {\bibfnamefont {S.}~\bibnamefont {Mizera}},\ and\ \bibinfo
  {author} {\bibfnamefont {J.}~\bibnamefont {Staunton}},\ }\href@noop {} {\
  (\bibinfo {year} {2025}{\natexlab{a}})},\ \Eprint
  {https://arxiv.org/abs/2510.25822} {arXiv:2510.25822 [hep-th]} \BibitemShut
  {NoStop}%
\bibitem [{\citenamefont {Chavda}\ \emph
  {et~al.}(2025{\natexlab{b}})\citenamefont {Chavda}, \citenamefont
  {McLoughlin}, \citenamefont {Mizera},\ and\ \citenamefont
  {Staunton}}]{Chavda:2025awr}%
  \BibitemOpen
  \bibfield  {author} {\bibinfo {author} {\bibfnamefont {A.}~\bibnamefont
  {Chavda}}, \bibinfo {author} {\bibfnamefont {D.}~\bibnamefont {McLoughlin}},
  \bibinfo {author} {\bibfnamefont {S.}~\bibnamefont {Mizera}},\ and\ \bibinfo
  {author} {\bibfnamefont {J.}~\bibnamefont {Staunton}},\ }\href@noop {} {\
  (\bibinfo {year} {2025}{\natexlab{b}})},\ \Eprint
  {https://arxiv.org/abs/2511.10613} {arXiv:2511.10613 [hep-th]} \BibitemShut
  {NoStop}%
\bibitem [{\citenamefont {Maldacena}(2003)}]{Maldacena:2002vr}%
  \BibitemOpen
  \bibfield  {author} {\bibinfo {author} {\bibfnamefont {J.~M.}\ \bibnamefont
  {Maldacena}},\ }\href {https://doi.org/10.1088/1126-6708/2003/05/013}
  {\bibfield  {journal} {\bibinfo  {journal} {JHEP}\ }\textbf {\bibinfo
  {volume} {05}},\ \bibinfo {pages} {013}},\ \Eprint
  {https://arxiv.org/abs/astro-ph/0210603} {arXiv:astro-ph/0210603}
  \BibitemShut {NoStop}%
\bibitem [{\citenamefont {Anninos}\ \emph {et~al.}(2015)\citenamefont
  {Anninos}, \citenamefont {Anous}, \citenamefont {Freedman},\ and\
  \citenamefont {Konstantinidis}}]{Anninos:2014lwa}%
  \BibitemOpen
  \bibfield  {author} {\bibinfo {author} {\bibfnamefont {D.}~\bibnamefont
  {Anninos}}, \bibinfo {author} {\bibfnamefont {T.}~\bibnamefont {Anous}},
  \bibinfo {author} {\bibfnamefont {D.~Z.}\ \bibnamefont {Freedman}},\ and\
  \bibinfo {author} {\bibfnamefont {G.}~\bibnamefont {Konstantinidis}},\ }\href
  {https://doi.org/10.1088/1475-7516/2015/11/048} {\bibfield  {journal}
  {\bibinfo  {journal} {JCAP}\ }\textbf {\bibinfo {volume} {11}},\ \bibinfo
  {pages} {048}},\ \Eprint {https://arxiv.org/abs/1406.5490} {arXiv:1406.5490
  [hep-th]} \BibitemShut {NoStop}%
\bibitem [{\citenamefont {Sleight}\ and\ \citenamefont
  {Taronna}(2021)}]{Sleight:2020obc}%
  \BibitemOpen
  \bibfield  {author} {\bibinfo {author} {\bibfnamefont {C.}~\bibnamefont
  {Sleight}}\ and\ \bibinfo {author} {\bibfnamefont {M.}~\bibnamefont
  {Taronna}},\ }\href {https://doi.org/10.1103/PhysRevD.104.L081902} {\bibfield
   {journal} {\bibinfo  {journal} {Phys. Rev. D}\ }\textbf {\bibinfo {volume}
  {104}},\ \bibinfo {pages} {L081902} (\bibinfo {year} {2021})},\ \Eprint
  {https://arxiv.org/abs/2007.09993} {arXiv:2007.09993 [hep-th]} \BibitemShut
  {NoStop}%
\bibitem [{\citenamefont {Harlow}\ and\ \citenamefont
  {Stanford}(2011)}]{Harlow:2011ke}%
  \BibitemOpen
  \bibfield  {author} {\bibinfo {author} {\bibfnamefont {D.}~\bibnamefont
  {Harlow}}\ and\ \bibinfo {author} {\bibfnamefont {D.}~\bibnamefont
  {Stanford}},\ }\href@noop {} {\  (\bibinfo {year} {2011})},\ \Eprint
  {https://arxiv.org/abs/1104.2621} {arXiv:1104.2621 [hep-th]} \BibitemShut
  {NoStop}%
\bibitem [{\citenamefont {Bzowski}\ \emph {et~al.}(2023)\citenamefont
  {Bzowski}, \citenamefont {McFadden},\ and\ \citenamefont
  {Skenderis}}]{Bzowski:2023nef}%
  \BibitemOpen
  \bibfield  {author} {\bibinfo {author} {\bibfnamefont {A.}~\bibnamefont
  {Bzowski}}, \bibinfo {author} {\bibfnamefont {P.}~\bibnamefont {McFadden}},\
  and\ \bibinfo {author} {\bibfnamefont {K.}~\bibnamefont {Skenderis}},\
  }\href@noop {} {\  (\bibinfo {year} {2023})},\ \Eprint
  {https://arxiv.org/abs/2312.17316} {arXiv:2312.17316 [hep-th]} \BibitemShut
  {NoStop}%
\bibitem [{\citenamefont {McFadden}\ and\ \citenamefont
  {Skenderis}(2010)}]{McFadden:2009fg}%
  \BibitemOpen
  \bibfield  {author} {\bibinfo {author} {\bibfnamefont {P.}~\bibnamefont
  {McFadden}}\ and\ \bibinfo {author} {\bibfnamefont {K.}~\bibnamefont
  {Skenderis}},\ }\href {https://doi.org/10.1103/PhysRevD.81.021301} {\bibfield
   {journal} {\bibinfo  {journal} {Phys. Rev. D}\ }\textbf {\bibinfo {volume}
  {81}},\ \bibinfo {pages} {021301} (\bibinfo {year} {2010})},\ \Eprint
  {https://arxiv.org/abs/0907.5542} {arXiv:0907.5542 [hep-th]} \BibitemShut
  {NoStop}%
\bibitem [{\citenamefont {Kontsevich}\ and\ \citenamefont
  {Segal}(2021)}]{Kontsevich:2021dmb}%
  \BibitemOpen
  \bibfield  {author} {\bibinfo {author} {\bibfnamefont {M.}~\bibnamefont
  {Kontsevich}}\ and\ \bibinfo {author} {\bibfnamefont {G.}~\bibnamefont
  {Segal}},\ }\href {https://doi.org/10.1093/qmath/haab027} {\bibfield
  {journal} {\bibinfo  {journal} {Quart. J. Math. Oxford Ser.}\ }\textbf
  {\bibinfo {volume} {72}},\ \bibinfo {pages} {673} (\bibinfo {year} {2021})},\
  \Eprint {https://arxiv.org/abs/2105.10161} {arXiv:2105.10161 [hep-th]}
  \BibitemShut {NoStop}%
\bibitem [{\citenamefont {Witten}(2021)}]{Witten:2021nzp}%
  \BibitemOpen
  \bibfield  {author} {\bibinfo {author} {\bibfnamefont {E.}~\bibnamefont
  {Witten}},\ }\href@noop {} {\  (\bibinfo {year} {2021})},\ \Eprint
  {https://arxiv.org/abs/2111.06514} {arXiv:2111.06514 [hep-th]} \BibitemShut
  {NoStop}%
\bibitem [{\citenamefont {Stefanyszyn}\ \emph {et~al.}(2024)\citenamefont
  {Stefanyszyn}, \citenamefont {Tong},\ and\ \citenamefont
  {Zhu}}]{Stefanyszyn:2024msm}%
  \BibitemOpen
  \bibfield  {author} {\bibinfo {author} {\bibfnamefont {D.}~\bibnamefont
  {Stefanyszyn}}, \bibinfo {author} {\bibfnamefont {X.}~\bibnamefont {Tong}},\
  and\ \bibinfo {author} {\bibfnamefont {Y.}~\bibnamefont {Zhu}},\ }\href
  {https://doi.org/10.1103/PhysRevLett.133.221501} {\bibfield  {journal}
  {\bibinfo  {journal} {Phys. Rev. Lett.}\ }\textbf {\bibinfo {volume} {133}},\
  \bibinfo {pages} {221501} (\bibinfo {year} {2024})},\ \Eprint
  {https://arxiv.org/abs/2406.00099} {arXiv:2406.00099 [hep-th]} \BibitemShut
  {NoStop}%
\bibitem [{\citenamefont {Salcedo}\ \emph {et~al.}(2023)\citenamefont
  {Salcedo}, \citenamefont {Lee}, \citenamefont {Melville},\ and\ \citenamefont
  {Pajer}}]{Salcedo:2022aal}%
  \BibitemOpen
  \bibfield  {author} {\bibinfo {author} {\bibfnamefont {S.~A.}\ \bibnamefont
  {Salcedo}}, \bibinfo {author} {\bibfnamefont {M.~H.~G.}\ \bibnamefont {Lee}},
  \bibinfo {author} {\bibfnamefont {S.}~\bibnamefont {Melville}},\ and\
  \bibinfo {author} {\bibfnamefont {E.}~\bibnamefont {Pajer}},\ }\href
  {https://doi.org/10.1007/JHEP06(2023)020} {\bibfield  {journal} {\bibinfo
  {journal} {JHEP}\ }\textbf {\bibinfo {volume} {06}},\ \bibinfo {pages}
  {020}},\ \Eprint {https://arxiv.org/abs/2212.08009} {arXiv:2212.08009
  [hep-th]} \BibitemShut {NoStop}%
\bibitem [{\citenamefont {Agui-Salcedo}\ and\ \citenamefont
  {Melville}(2023)}]{Agui-Salcedo:2023wlq}%
  \BibitemOpen
  \bibfield  {author} {\bibinfo {author} {\bibfnamefont {S.}~\bibnamefont
  {Agui-Salcedo}}\ and\ \bibinfo {author} {\bibfnamefont {S.}~\bibnamefont
  {Melville}},\ }\href {https://doi.org/10.1007/JHEP12(2023)076} {\bibfield
  {journal} {\bibinfo  {journal} {JHEP}\ }\textbf {\bibinfo {volume} {12}},\
  \bibinfo {pages} {076}},\ \Eprint {https://arxiv.org/abs/2308.00680}
  {arXiv:2308.00680 [hep-th]} \BibitemShut {NoStop}%
\bibitem [{\citenamefont {Goodhew}\ \emph
  {et~al.}(2021{\natexlab{a}})\citenamefont {Goodhew}, \citenamefont
  {Jazayeri},\ and\ \citenamefont {Pajer}}]{Goodhew:2020hob}%
  \BibitemOpen
  \bibfield  {author} {\bibinfo {author} {\bibfnamefont {H.}~\bibnamefont
  {Goodhew}}, \bibinfo {author} {\bibfnamefont {S.}~\bibnamefont {Jazayeri}},\
  and\ \bibinfo {author} {\bibfnamefont {E.}~\bibnamefont {Pajer}},\ }\href
  {https://doi.org/10.1088/1475-7516/2021/04/021} {\bibfield  {journal}
  {\bibinfo  {journal} {JCAP}\ }\textbf {\bibinfo {volume} {04}},\ \bibinfo
  {pages} {021}},\ \Eprint {https://arxiv.org/abs/2009.02898} {arXiv:2009.02898
  [hep-th]} \BibitemShut {NoStop}%
\bibitem [{\citenamefont {Goodhew}\ \emph
  {et~al.}(2021{\natexlab{b}})\citenamefont {Goodhew}, \citenamefont
  {Jazayeri}, \citenamefont {Lee},\ and\ \citenamefont
  {Pajer}}]{Goodhew:2021oqg}%
  \BibitemOpen
  \bibfield  {author} {\bibinfo {author} {\bibfnamefont {H.}~\bibnamefont
  {Goodhew}}, \bibinfo {author} {\bibfnamefont {S.}~\bibnamefont {Jazayeri}},
  \bibinfo {author} {\bibfnamefont {M.~H.~G.}\ \bibnamefont {Lee}},\ and\
  \bibinfo {author} {\bibfnamefont {E.}~\bibnamefont {Pajer}},\ }\href
  {https://doi.org/10.1088/1475-7516/2021/08/003} {\bibfield  {journal}
  {\bibinfo  {journal} {JCAP}\ }\textbf {\bibinfo {volume} {08}},\ \bibinfo
  {pages} {003}},\ \Eprint {https://arxiv.org/abs/2104.06587} {arXiv:2104.06587
  [hep-th]} \BibitemShut {NoStop}%
\end{thebibliography}%

\end{document}